\documentclass[referee]{raa}
\usepackage{graphicx,times}
\usepackage{natbib}
\usepackage{amssymb,amsmath}
\bibpunct{(}{)}{;}{a}{}{,}

\usepackage[a4paper=true,dvipdfm=true,pagebackref=true]{hyperref}
\hypersetup{pdftitle = The title of my PDF, pdfauthor = My name, pdfsubject= The subject, pdfkeywords = keyword1 keyword2 keyword3}
\hypersetup{colorlinks = true, linkcolor = green, anchorcolor = red, citecolor = blue, filecolor = red, pagecolor = red, urlcolor = red}
\input{psfig.sty}

\begin{document}

\title{Estimations of the cosmological parameters from the observational variation of the fine structure constant
} 

 \volnopage{ {\bf 2012} Vol.\ {\bf X} No. {\bf XX}, 000--000}
   \setcounter{page}{1}

\author{Zhong-Xu Zhai\inst{1,2}
        Xian-Ming Liu\inst{3,2}
        Zhi-Song Zhang\inst{4}
        Tong-Jie Zhang\inst{1,5} 
}

\institute{Department of Astronomy, Beijing Normal University, Beijing, 100875, China \label{a}; {\it tjzhang@bnu.edu.cn}
           \and
           Department of Physics, Institute of Theoretical Physics, Beijing Normal University, Beijing, 100875, China \label{b}
           \and
           Department of Physics, Hubei University for Nationalities, Enshi Hubei, 445000, China \label{c}
           \and
           Department of Aerospace Engineering, School of Astronautics, Harbin Institute of Technology (HIT), Harbin Heilongjiang, 150001, China \label{d}
           \and
           Center for High Energy Physics, Peking University, Beijing, 100871, China \label{e}
\vs \no \\
   {\small Received XXXX; accepted XXXX}
}

\abstract{
We present the constraints on the Quintessence scalar field model
from the observational data of the variation of the fine structure
constant obtained from Keck and VLT telescopes. Within the
theoretical frame proposed by (\cite{Bekenstein..alpha}),
the constraints on the parameters of the Quintessence scalar field
model are obtained. By the consideration of the prior of
$\Omega_{m0}$ as WMAP 7 suggests (\cite{Komatsu..CMB}), we obtain
various results of the different samples. Based on these results, we
also calculate the probability density function of the coupling
constant $\zeta$. The best-fit values show a consistent relationship
between $\zeta$ and the different experimental results. In our work,
we test two different potential models, namely, the inverse power
law potential and the exponential potential. The results show that
both the large value of the parameters in the potential and the
strong coupling can cause the variation of fine structure constant.
\keywords{quasars: general --- scalar field --- cosmological constraint}
}
   \authorrunning{Z.-X. Zhai et al. }            
   \titlerunning{Estimations of the cosmological parameters}  
   \maketitle

\section{Introduction} \label{sec:introduction}

Fundamental constants play important roles in physics and its
mathematical laws. By the information they contain, one can describe
the phenomena of nature and obtain a better understanding of the
real world (\cite{Uzan..review}). However, one may suspect whether the
constants are real "constants", i.e. do the constants vary
 with time or space? This question was probably first asked
by Dirac with his famous "Large Numbers Hypothesis"(LNH)
(\cite{Dirac..1,Dirac..2}). Thereafter, several works have been done
to investigate the underground principle, including researches of
the variations of the constants and the measurements of their
precise values. We refer the readers to the reviews
Ref.  (\cite{Barrow..review,Barrow..review2,Chiba..review,Damour..review,Flambaum..review,Garcia-Berro..review,Karshenboim..review,Uzan..review2003})
for more comprehensive discussions.

In order to unify the fundamental interactions theoretically,
 different theories are proposed including string derived field theories,
brane-world theories, Kaluza-Klein theories which are based on the
introduction of the extra dimensions. Among the constants, the fine
structure constant $\alpha$ which measures the strength of the
electromagnetic interaction attracts a lot of attention. In 1982,
Bekenstein  proposed a different theoretical framework to study
$\alpha$ variability where a linear coupling between a scalar field
and the electromagnetic field was introduced
(\cite{Bekenstein..alpha}). This theory satisfies the general
conditions: covariance, gauge invariance, causality, and time-
reversal invariance of electromagnetism. Later on, this proposal was
generalized and improved by \cite{SBM..2002}. So far we have several
different theories which can describe the time evolution of the
gauge coupling constants. However, whether the theoretical
predictions can provide consistent results with the experimental
ones should be asked. In this paper, we limit ourselves to study the
Bekenstein model and the time-evolving behavior of the fine
structure constant. The experiments which imply a time-related
$\alpha$ include the observation of the Oklo natural nuclear reactor
(\cite{Damour..oklo,Olive..oklo}), Big Bang
Nuclesynthesis(BBN) (\cite{Avelino..BBN,Martins..BBN,Nollett..BBN}),
Cosmic Microwave Background (CMB) measurements
(\cite{Avelino..BBN,Martins..BBN,Landau..CMB,Menegoni..CMB,Nakashima..CMB}),
absorption spectra of distant Quasars(QSOs)
(\cite{Chand..QSO,Murphy..QSO,Srianand..QSO,Webb..QSO}), and so on.
These observations give different measurements of $\alpha$ at
different cosmological evolution periods. Among these observations,
QSO absorption lines provide a powerful probe of the variation of
$\alpha$ and a large data sample. The methods studying this
observation results include the \textit{alkali doublet method} (AD),
the \textit{many-multiplet method} (MM), the \textit{revised
many-multiplet method} (RMM), and the \textit{Single ion
differential alpha measurement method} (SIDAM) (\cite{Uzan..review}).
Because the observation by MM gives the widest range of redshift
($0.22<z<4.2$) (\cite{Murphy..QSO,Landau..QAO}), it may contain more
information of the cosmological evolution than the others. Thus we
will mainly focus on these measurement in the present work. More
details about the observational data will be presented in
Sec.\ref{data}

On the other hand, since its discovery more than ten years ago, the
cosmic accelerated expansion has been demonstrated by the
observations of type Ia supernovae and this phenomena is accepted
widely
(\cite{Eisenstein..BAO,Hicken..Sne,Komatsu..CMB,Percival..BAO,Riess..Sne,Spergel..CMB}).
In order to explain this amazing discovery, a great variety of
attempts have been done including the introduction of dark energy
and the modified gravity theories (\cite{Tsujikawa..dark}). Among
these proposals, the scalar field as a dynamical dark energy model
was studied widely and deeply
(\cite{Chen..scalar,Li..scalar,Samushia..scalar}). Therefore, the
cosmological variation of $\alpha$ induced by coupling with the
Quintessence, which is a typical scalar field dark energy,
 is worth studying in order to find
if the QSO observations contain the information of the cosmic
accelerated expansion. In other words, whether the QSO observations
can give a consistent result with other cosmological probes, such as
Type Ia Supernovae (SNe Ia), CMB, Baryon acoustic oscillation (BAO),
Observational Hubble parameter Data (OHD)
(\cite{Ma..OHD,Moresco..OHD,Zhang..OHD}) and so forth, should be
tested. Moreover, if the observation of QSO absorption lines
provides consistent results with the ones listed above, can it be
thought as an indirect proof of the existence of the scalar field
(Quintessence)? This will be a very interesting question. In
addition, we should notice that there is some difference between the
QSO observations in Ref. \cite{Murphy..QSO,Webb..QSO} and
Ref. \cite{Chand..QSO,Srianand..QSO}. The results analyzed by these two
MM methods show an inconsistency of the time evolution of $\alpha$.
Thus what information of cosmological evolution these data contained
respectively should be studied.

Following this direction, we constrain the cosmological parameters
of the Quintessence dark energy model with the variational $\alpha$
data from the observation of the QSO absorption lines. One should
note that there are several freedoms in choosing the form of the
scalar field potential which plays an important role in the scalar
field evolution. In our paper, we firstly focus on the inverse
power-law potential $V(\phi)\propto\phi^{-n}$, where $n$ is a
nonnegative constant (\cite{Peebles..Ratra,Ratra..Peebles}). This
assumption has several advantages such as it can reduce to the
standard $\Lambda$CDM case when $n=0$ and contain the solutions
which can alleviate the fine-tuning problem (\cite{Watson..inverse}).
Recent researches of the mass scale of the inverse power law
potential show that the field value at present is of order the
Planck mass ($\phi_{0}\sim M_{P}$)
(\cite{Tsujikawa..dark,Steinhardt..mass,Zlatev..mass}). For
comparison, we also consider another potential model $V(\phi)\propto
e^{-\lambda\phi}$, where $\lambda$ is a positive constant
(\cite{Ratra..Peebles}). This model was first motivated by the anomaly
of the dilatation symmetry in the particle physics and has the
tracker solution at the late time
(\cite{Wetterich..exponen,Doran..exponen}). In this paper we just
consider a spatially-flat Quintessence model.

Many previous works that constrain the parameters of the
Quintessence dark energy model show that the universe is composed by
about 30$\%$ nonrelativistic matter while the dark energy
contributes nearly 70$\%$. And the parameter $n$ (of the inverse
power law potential) and $\lambda$ (of the exponential potential)
which affects the evolution behavior of the scalar field directly
both favor small values
(\cite{Samushia..scalar,Bozek..exponen,Wang..exponen}). So we should
ask to what extent are the constraints from QSO observations
consistent with these results. The possibility of studying the fine
structure constant under the dark energy models has been proposed
from various aspects, including the reconstruction of the dark
energy equation of state
(\cite{Avelino..recon,Nunes..recon,Parkinson..recon}) or combined with
other cosmological observations (\cite{Amendola..sne}). In this paper
we will discuss the possibility of constraining the quintessence
dark energy model with the direct measurements of the variation of
fine structure constant.

Our paper is organized as follows. In Sec.\ref{basic}, we will
present the basic formulas of the Quintessence-$\alpha$ model. The
data used and the corresponding constraints are shown in
Sec.\ref{data}. The conclusion is presented in Sec.\ref{conclusion}.

\section{Quintessence and the electromagnetic couplings}\label{basic}

We consider a spatially-flat FRW cosmology where the metric can be
written as
\begin{equation}
ds^{2}=-dt^{2}+a(t)^{2}(dr^{2}+r^{2}d\theta^{2}+r^{2}\sin^{2}\theta
d\varphi^{2}),
\end{equation}
where $a$ is the scale factor. Under this geometrical background,
the evolution of the Quintessence scalar field $\phi$ is determined
by the Friedmann equation and the Klein-Gordon equation
\begin{equation}\label{Eq:Friedmann}
H^{2}=(\frac{\dot{a}}{a})^{2}=\frac{8\pi}{3M_{p}^{2}}\sum \rho_{i},
\end{equation}
\begin{equation}\label{Eq:KG}
  \ddot{\phi}+3H\dot{\phi}+\frac{dV}{d\phi}=0,
\end{equation}
where $M_{p}$ is the Planck mass, the overdot is the derivative with
respect to the cosmic time $t$, $\rho$ stands for the density and
$i=m,\phi$ runs over the matter (including dark matter) and scalar
components. The relevant equations of state are $\omega_{m}=0$ for
matter and $\omega_{\phi}=p_{\phi}/\rho_{\phi}$ for scalar field
where
\begin{equation}
  p_{\phi}=\frac{\dot{\phi}^{2}}{2}-V(\phi),\quad \rho_{\phi}=\frac{\dot{\phi}^{2}}{2}+V(\phi).
\end{equation}
In our calculation, the function form of the potential are
\begin{eqnarray}
Model I \qquad  & V(\phi)=\kappa M_{p}^{2}\phi^{-n},  \nonumber\\
Model II \qquad & V(\phi)=V_{0}e^{-\lambda\phi}, \label{Eq:V}
\end{eqnarray}
where $\kappa$, $V_{0}$ are non-negative constants, $n$ and
$\lambda$ are the parameter will be constrained by the data. These
kinds of scalar field model was first studied by Peebles and Ratra
in 1988 and further explored especially in explaining the dark
energy
problem (\cite{Chen..scalar,Samushia..scalar,Russo..exponen,Binetruy1..exponen,Binetruy2..exponen,Ferreira..exponen})(and
references therein). By the definitions of the dimensionless
parameters
\begin{equation}
  \Omega_{m}=\frac{8\pi\rho_{m}}{3M_{p}^{2}H^{2}}=\frac{\rho_{m}}{\rho_{m}+\rho_{\phi}},\quad \Omega_{\phi}=\frac{8\pi\rho_{\phi}}{3M_{p}^{2}H^{2}}=\frac{\rho_{\phi}}{\rho_{m}+\rho_{\phi}},
\end{equation}
the Friedmann Equation Eq.(\ref{Eq:Friedmann}) can be rewritten in a
simple form
\begin{equation}
  \Omega_{m}+\Omega_{\phi}=1.
\end{equation}
So far our model is determined by only two parameters
($\Omega_{m0},n$) for Model I and ($\Omega_{m0},\lambda$) for Model
II, where the subscript 0 stands for the present value. This
parameter set is the key point that will be constrained by the
observational data.

Considering an interaction between a Quintessence field $\phi$ and
an electromagnetic field $F_{\mu\nu}$, we can write its Lagrangian
density as
\begin{equation} \label{Eq:lag}
  \mathcal {L}_{F}(\phi)=-\frac{1}{4}B_{F}(\phi)F_{\mu\nu}F^{\mu\nu},
\end{equation}
where $B_{F}(\phi)$ is the function that describes the coupling
behavior. One should note that the addition of this interaction term
does not affect the evolution of the quintessence scalar field. This
is due to the fact that the statistical average of
$F_{\mu\nu}F^{\mu\nu}$ over a current state of the universe is zero
(\cite{Copeland..zeta,Marra..BF}). Thus Eq.(\ref{Eq:KG}) is still
applicable. The Lagrangian form Eq.(\ref{Eq:lag}) allows us to
define a new "effective" fine structure constant
\begin{equation}
  \alpha(\phi)=\frac{\alpha_{0}}{B_{F}(\phi)},
\end{equation}
where $\alpha_{0}$ is the current value. By the use of this equation
we can obtain a relative variation of $\alpha$
\begin{equation}\label{Eq:alphath}
  \frac{\Delta\alpha}{\alpha}=\frac{\alpha(\phi)-\alpha_{0}}{\alpha_{0}}=\frac{1-B_{F}(\phi)}{B_{F}(\phi)}.
\end{equation}
Apparently, the evolution of $\alpha$ is directly affected by $\phi$
and the functional form $B_{F}(\phi)$. From the theoretical view,
there are many choices in defining $B_{F}$ which leads to different
$\alpha$ behaviors. The authors of Ref. (\cite{Marra..BF}) give a detailed
discussion about $B_{F}(\phi)$ which contains many different cases.
In our paper, we will consider the simplest case which is a linear
form and corresponds to the original Bekenstein proposal
(\cite{Bekenstein..alpha}),
\begin{equation}\label{Eq:BF}
  B_{F}(\phi)=1-\zeta(\phi-\phi_{0}),
\end{equation}
where the constant $\zeta$ describes the strength of the coupling
between the scalar field and the electromagnetic field. We will see
that the parameter sets $(\Omega_{m0},n,\zeta)$ and
$(\Omega_{m0},\lambda,\zeta)$ completely describe the evolution
behavior of the Quintessence-$\alpha$ Model I and Model II
respectively.

\section{the observational QSO data and constraints}\label{data}

\begin{figure}[htb]
\centerline
{\psfig{figure=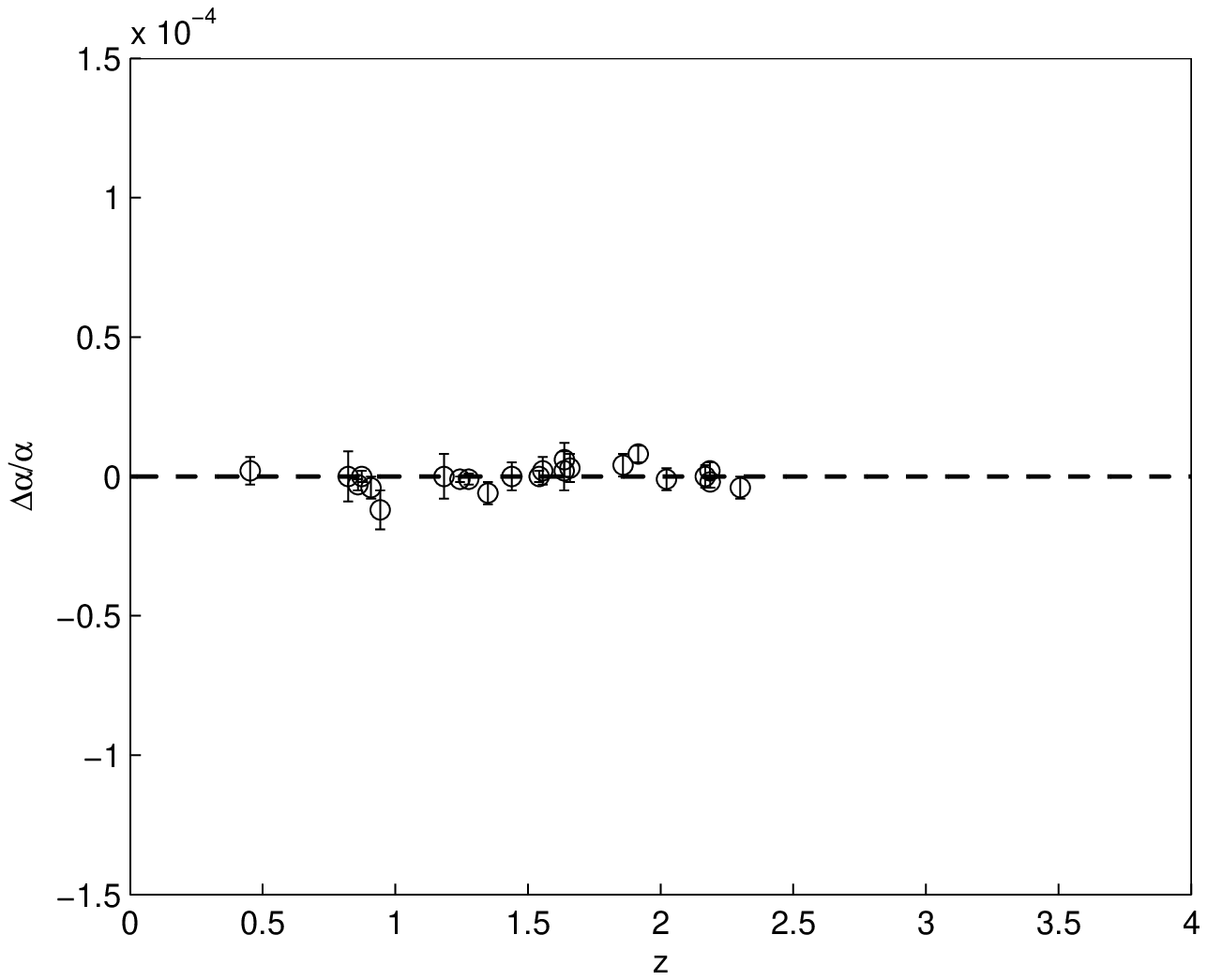,width=2.1truein,height=1.8truein}
\psfig{figure=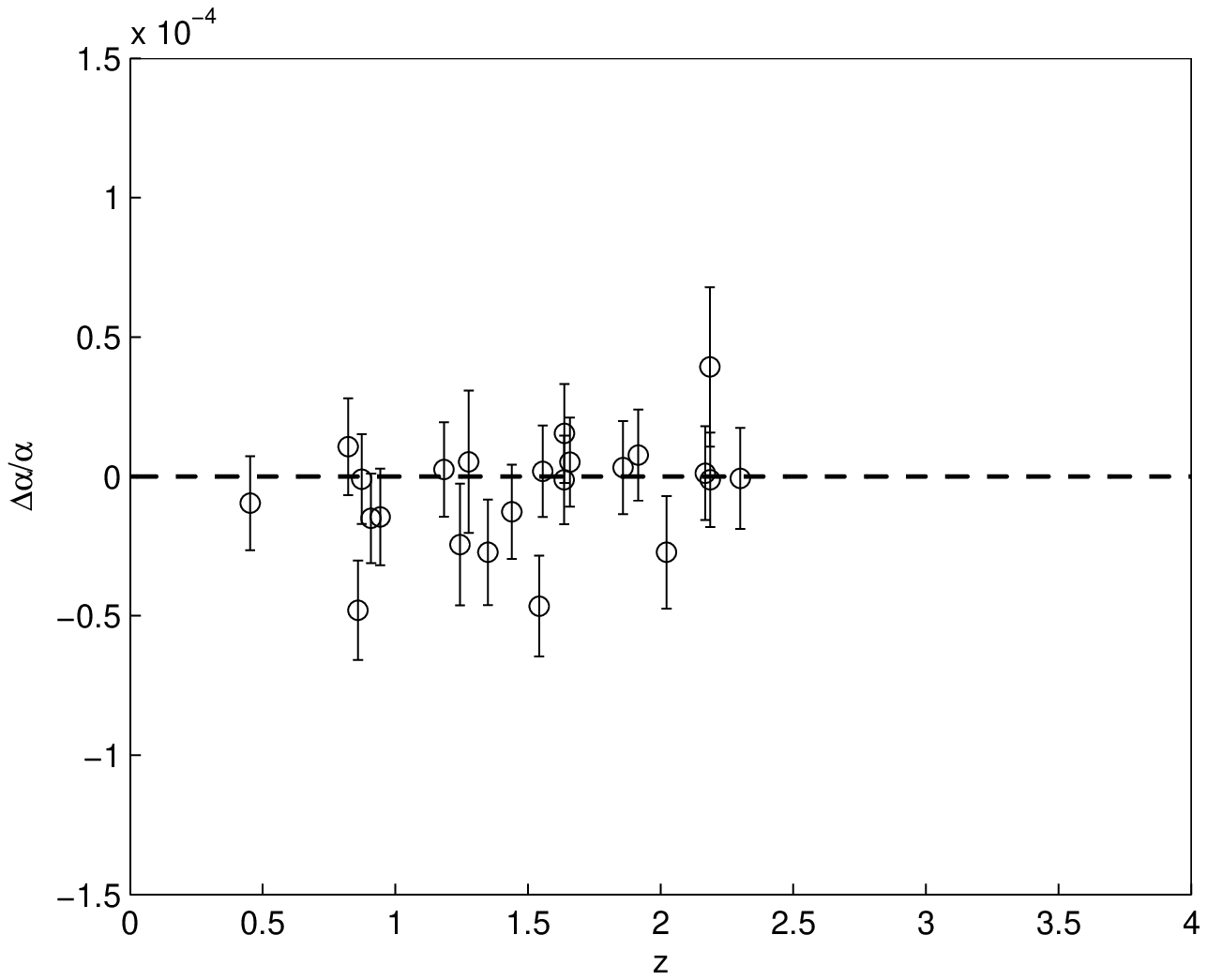,width=2.1truein,height=1.8truein}
\psfig{figure=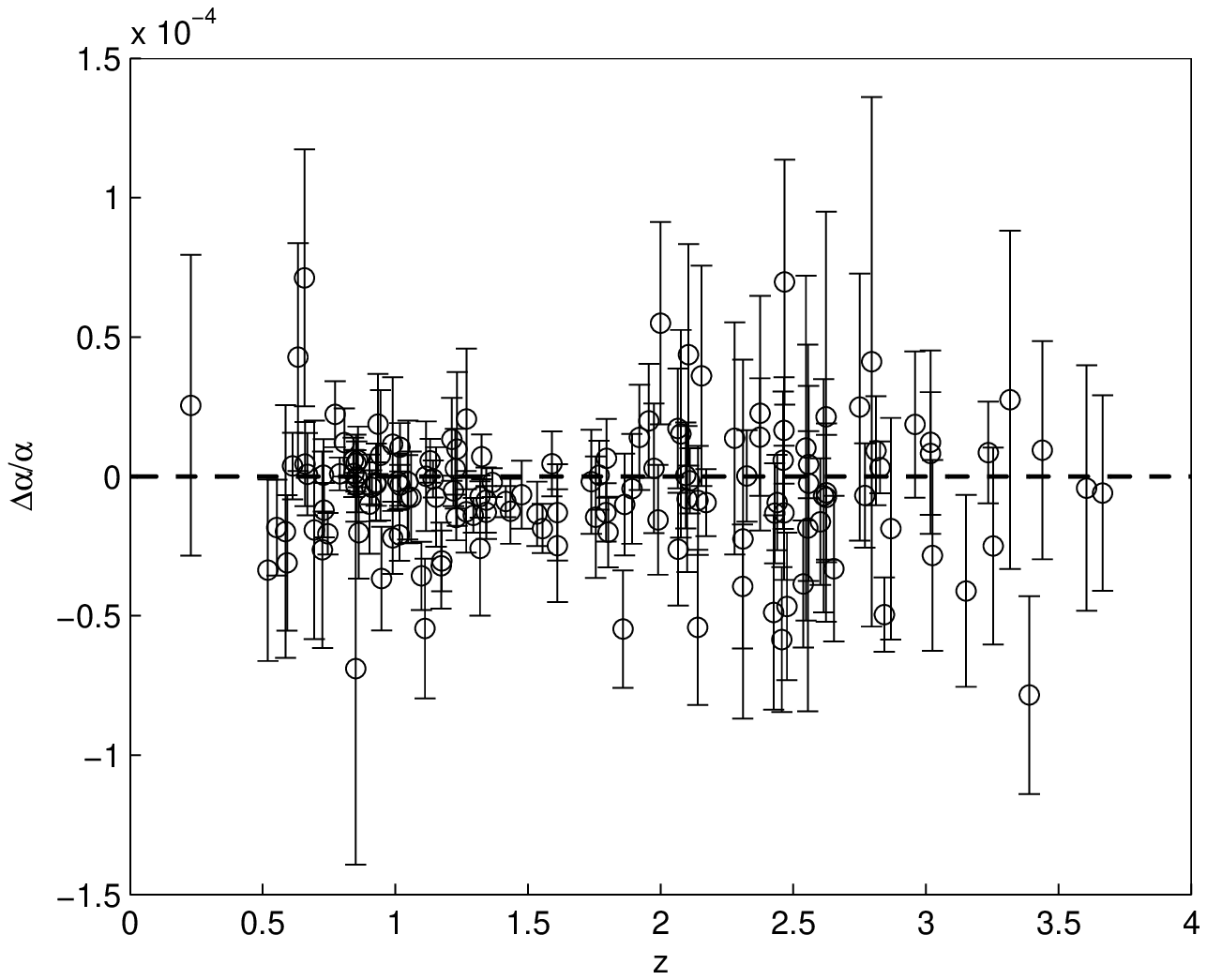,width=2.1truein,height=1.8truein}
\hskip 0.01in} \caption{The direct measurements of
$\Delta\alpha/\alpha$ with respect to the redshift $z$: VCS23(left);
VMW23(middle); KMW128(right). The dashed curve is the horizonal line
indicating no variation.} \label{fig:data}
\end{figure}

\subsection{the observational QSO data}
The MM method as a generalization of the AD method was first
proposed in Ref. \cite{Dzuba..MM}. It was first applied in Ref.
(\cite{Webb..QSO}) to analyze the distant QSO absorption lines
observed by Keck which is located in Hawaii. Their result shows a
variation of $\alpha$ in the redshift range of $0.6<z<1.6$. Later
on, more QSO systems were observed and the data sample was enlarged.
The updated results which are based on a statistical analysis
including 143 absorption systems show that
$\Delta\alpha/\alpha=(-0.57\pm0.11)\times10^{-5}$ in the redshift
range of $0.2<z<4.2$ (\cite{Murphy..QSO}). We will use this data
sample to test the Quintessence-$\alpha$ model. For convenience, we
use "KWM143" as an abbreviation for this sample. Although there are
some differences in analyzing the low-z and high-z absorption
systems, we will combine the total 143 data to do the calculation
and neglect the tiny discrepancies.

On the other hand, a further independent statistical study was
completed in Ref. \cite{Chand..QSO,Srianand..QSO} based on the
observations of VLT. Their calculation favors a different result of
$\Delta\alpha/\alpha=(-0.06\pm0.06)\times10^{-5}$  which shows a
nearly unchanged $\alpha$ in the redshift range of $0.4<z<2.3$.
However, this analysis was challenged by \cite{Murphy..QSO2} who
used the same reduced data and got a result of
$\Delta\alpha/\alpha=(-0.44\pm0.16)\times10^{-5}$ in the same
redshift range. However this result is not reliable because of its
larger value of the reduced $\chi^{2}$ in Ref. \cite{Murphy..QSO2}.
Therefore it is necessary to consider the additional scatter and the
systematic error which derive the most conservative weighted mean
result becomes $\Delta\alpha/\alpha=(-0.64\pm0.36)\times10^{-5}$.
This result also prefers a non-zero variation of the fine structure
constant. Since then, this contradictory results were discussed
several times by different research groups and the principles behind
the observations were explored from many different aspects
(\cite{King..QSO,Barrow..alpha,Bento..alpha,Bisabr..fr,Calabrese..alpha,Fujii..alpha,Gutierrez..alpha,Lee..alpha,Mosquera..alpha,Tedesco,Toms..quantum,Avelino..re..alpha,Farajollahi..alpha,Martinelli..alpha,Thompson..alpha}).
Recently, an intensive debate in literature was proposed by
\cite{Berengut..spatial,Berengut..alpha,Webb..spatial}. They propose
that the observed spatial variation of $\alpha$ is really not an
artificial effect, resulting from the fact that Keck and VLT are
located at different hemispheres. Therefore perhaps it is worth
noticing to study the possibility of the spatial variation of
$\alpha$ in the quintessence model, but the new physics may be taken
into account. And these will be the future focus of our researches.
In the present calculation, we will use these two different
data samples independently and constrain the Quintessence-$\alpha$
model respectively. One of our goals is to explore the reasons that
cause the above two different results, i.e. aiming at finding out
whether the coupling strength or the cosmological evolution of
quintessence leads to the discrepancy between them. In the
following, we use "VCS23" and "VWM23" as abbreviations for these two
samples. In FIG.\ref{fig:data}, we
plot the direct measurements of $\Delta\alpha/\alpha$ of these three data samples listed above.

From Eq.(\ref{Eq:alphath}) and (\ref{Eq:BF}), we can see that the
value of $\zeta$ effects the evolution of $\Delta\alpha/\alpha$
directly. From the tests of the equivalence principle the coupling
is constrained to be $|\zeta|<10^{-3}$
(\cite{Copeland..alpha,Olive..zeta}). Furthermore,
\cite{Copeland..zeta} used a simple estimation to obtain an
approximate value of $\zeta\approx10^{-5}$ which is under the
assumption of inverse power law potential and the QSO observations.
In our paper, we consider $\zeta$ as a free parameter to be
constrained by the data and compare the results with the previous
works such as the Equivalence Principle test
(\cite{Avelino..zeta,Damour..zeta,Will..zeta}) which shows that
$|\zeta|<5\times10^{-4}$.

\subsection{Constraints with a prior of $\Omega_{m0}$} \label{sec:gaussian}

\begin{figure*}
\centerline
{\psfig{figure=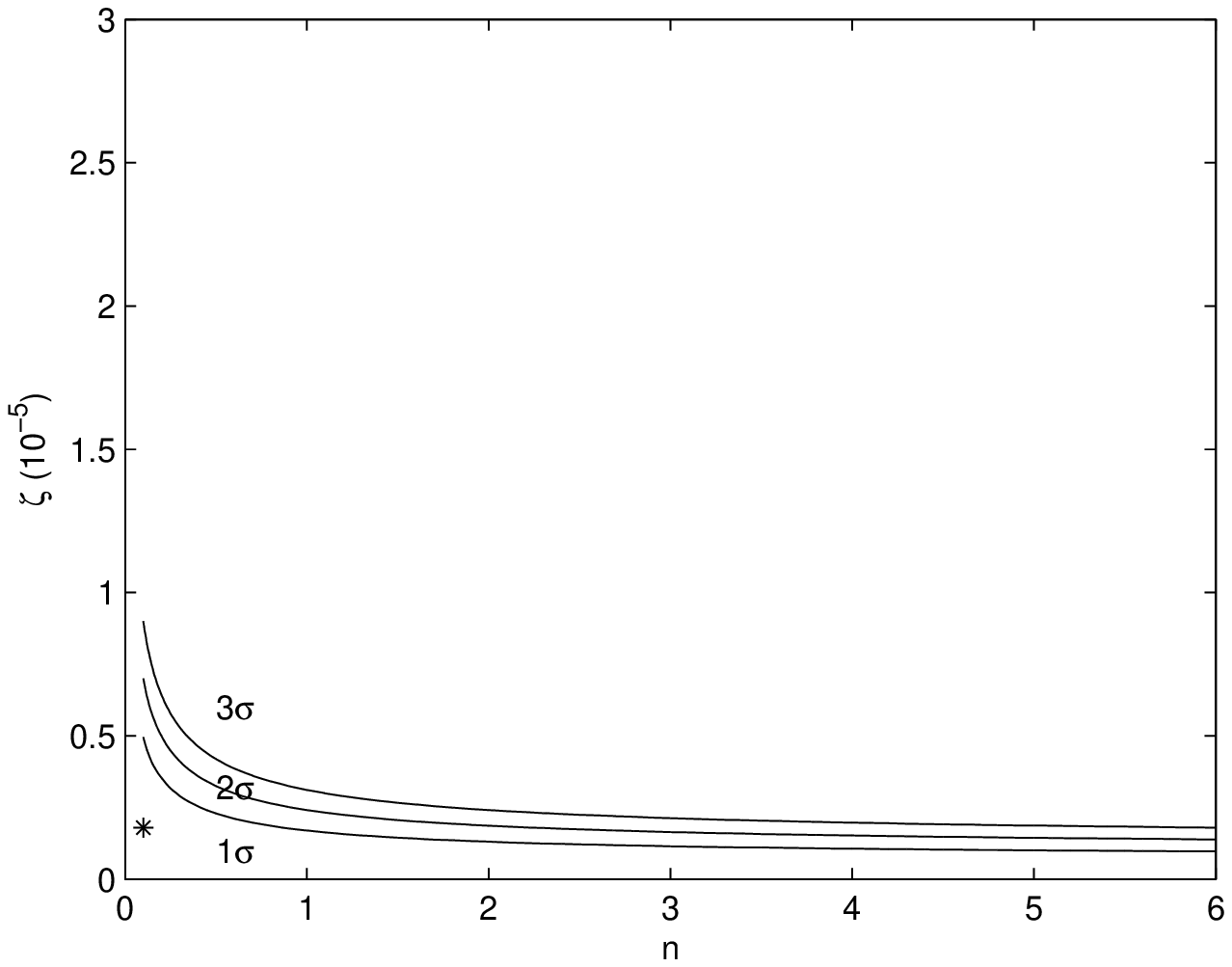,width=3.2truein,height=2.7truein}
 \psfig{figure=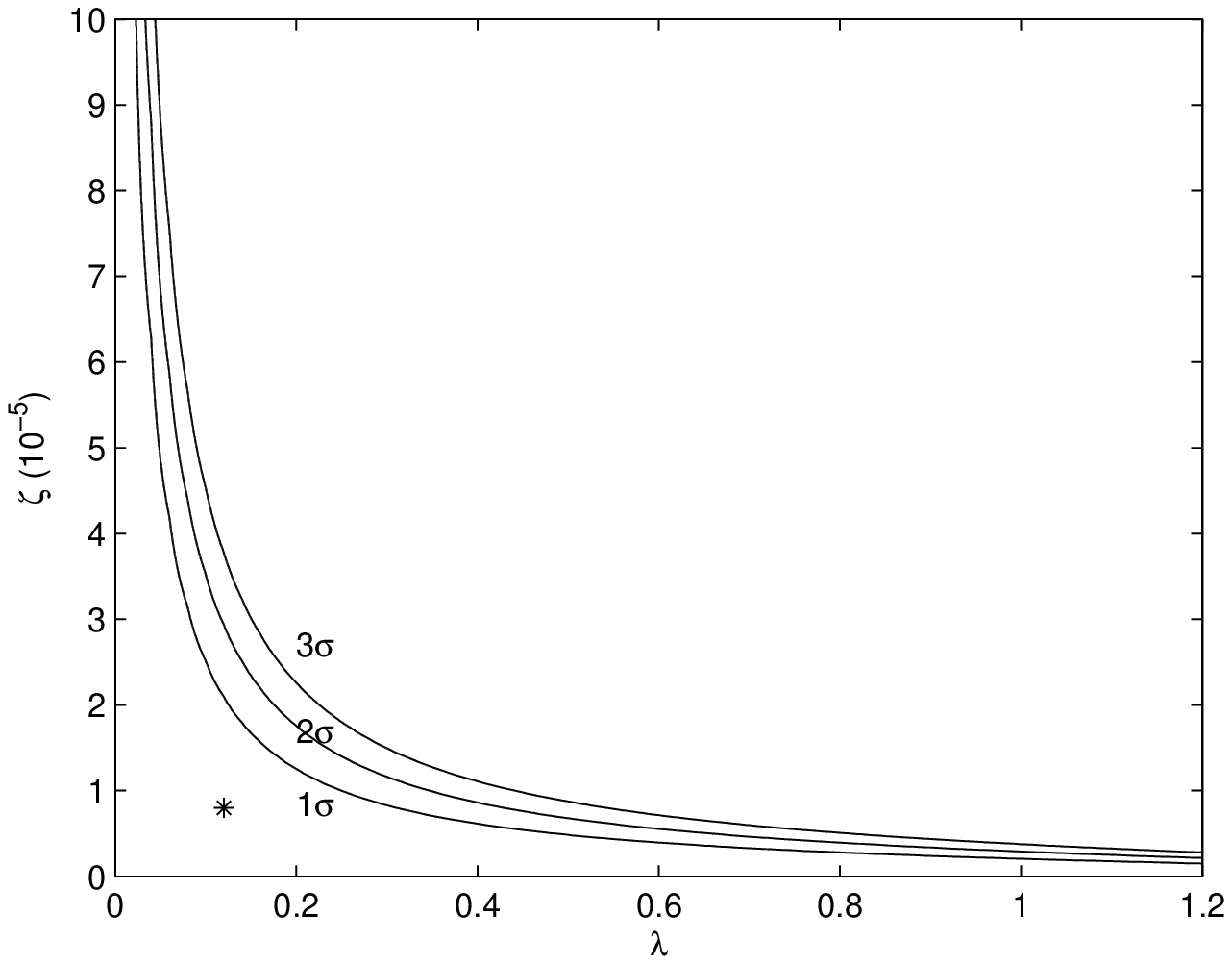,width=3.2truein,height=2.7truein}
\hskip 0.01in} \caption{$Left (Model I)$: The confident regions of
$(n,\zeta)$ obtained from VCS23 with a prior of
$\Omega_{m0}=0.275\pm0.016$. The best-fit results which is indicated
by the star are $(n,\zeta)=(0.1, 0.18\times10^{-5})$ with
$\chi^{2}_{min}=27.7919$. $Right (Model II)$: The confident regions
of $(\lambda,\zeta)$ obtained from VCS23 with a prior of
$\Omega_{m}$. The best-fit results which is indicated by the star
are $(\lambda,\zeta)=(0.12, 0.80\times10^{-5})$ with
$\chi^{2}_{min}=27.6141$.} \label{fig:VCS23_Om}
\end{figure*}

\begin{figure*}
\centerline
{\psfig{figure=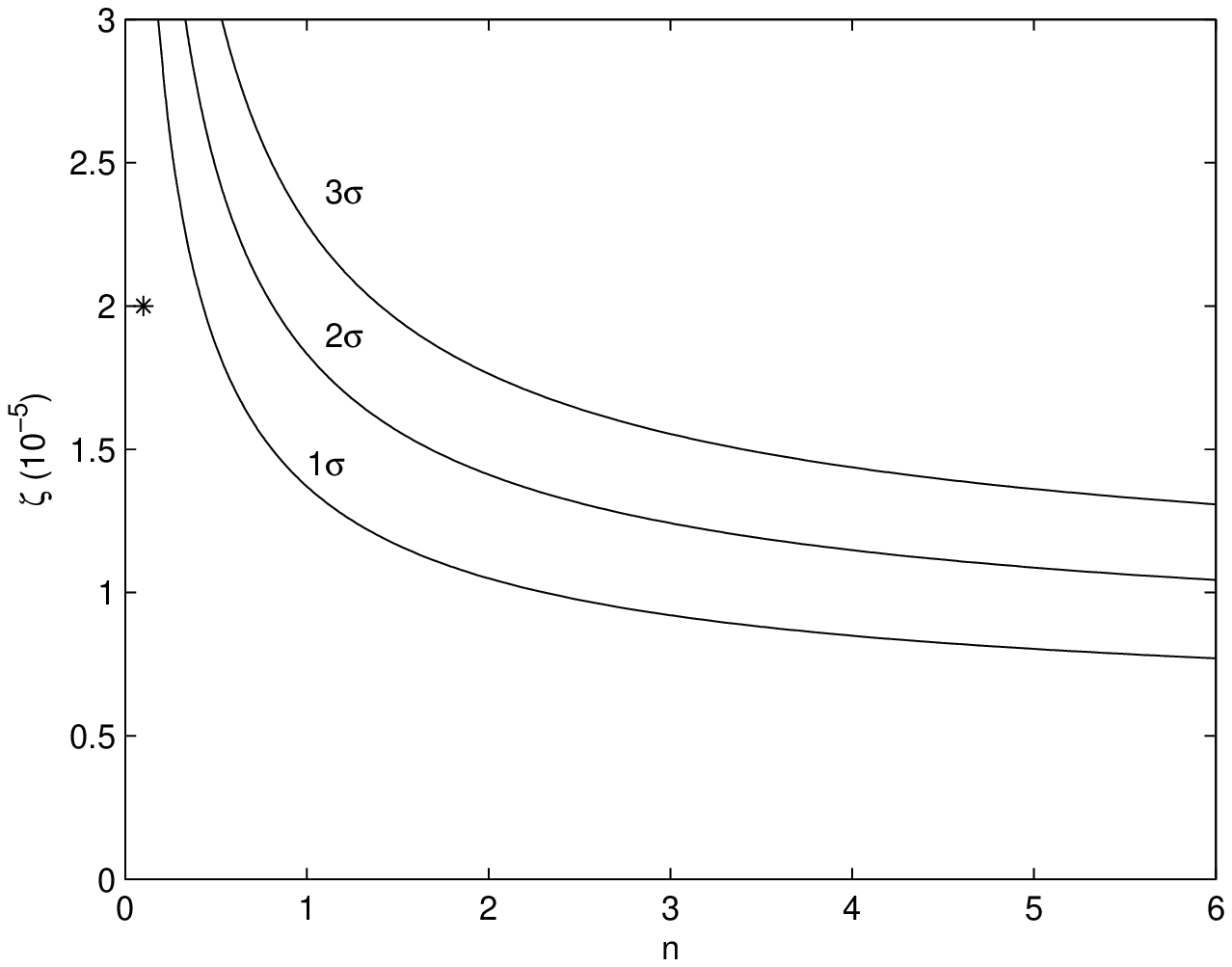,width=3.2truein,height=2.7truein}
 \psfig{figure=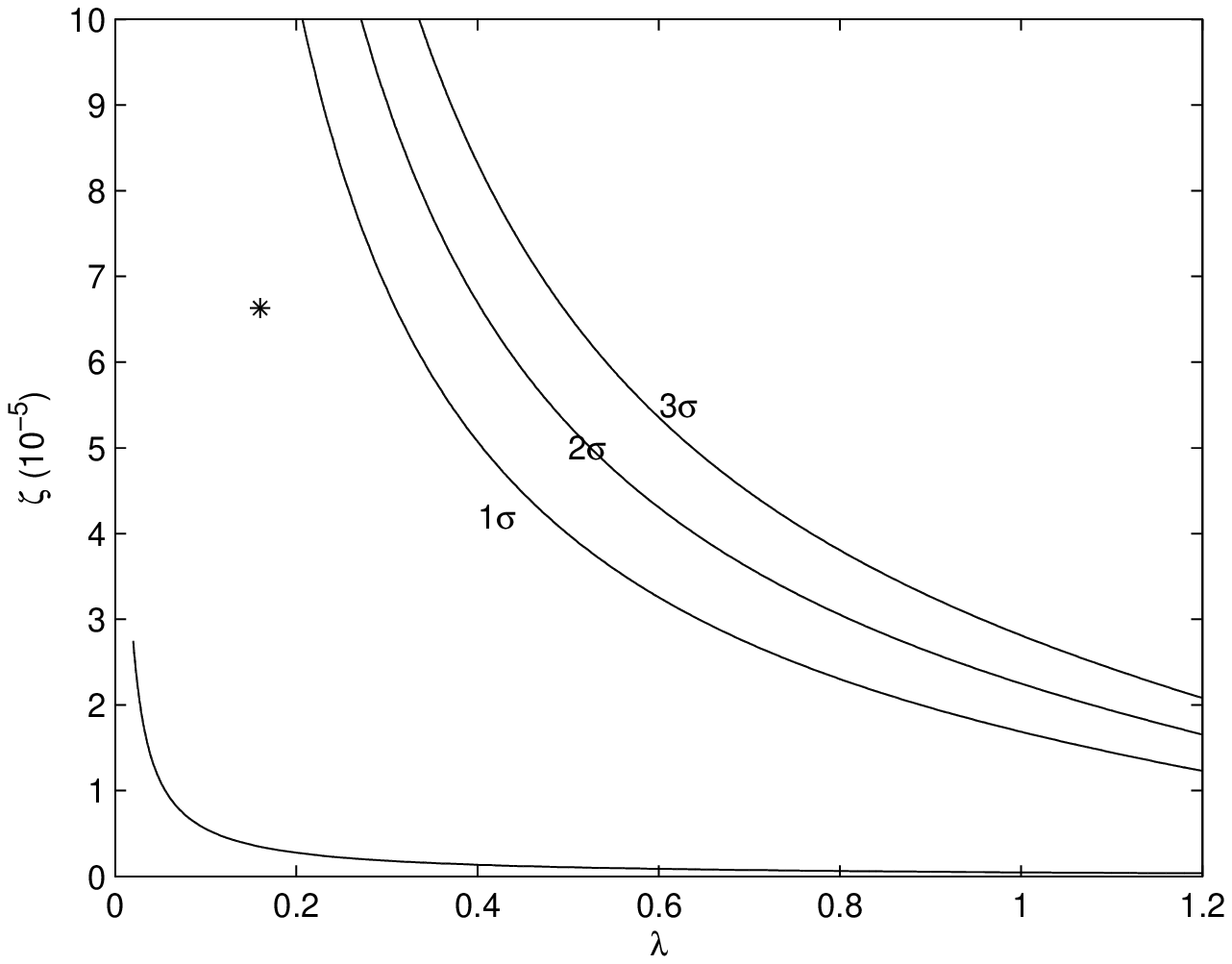,width=3.2truein,height=2.7truein}
\hskip 0.01in} \caption{$Left (Model I)$: The confident regions of
$(n,\zeta)$ obtained from VWM23 with a prior of
$\Omega_{m0}=0.275\pm0.016$. The best-fit results which is indicated
by the star are $(n,\zeta)=(0.1, 2.0\times10^{-5})$ with
$\chi^{2}_{min}=29.2648$. $Right (Model II)$: The confident regions
of $(\lambda,\zeta)$ obtained from VWM23 with a prior of
$\Omega_{m}$. The best-fit results which is indicated by the star
are $(\lambda,\zeta)=(0.16, 6.63\times10^{-5})$ with
$\chi^{2}_{min}=28.8861$.} \label{fig:VWM23_Om}
\end{figure*}

\begin{figure*}
\centerline
{\psfig{figure=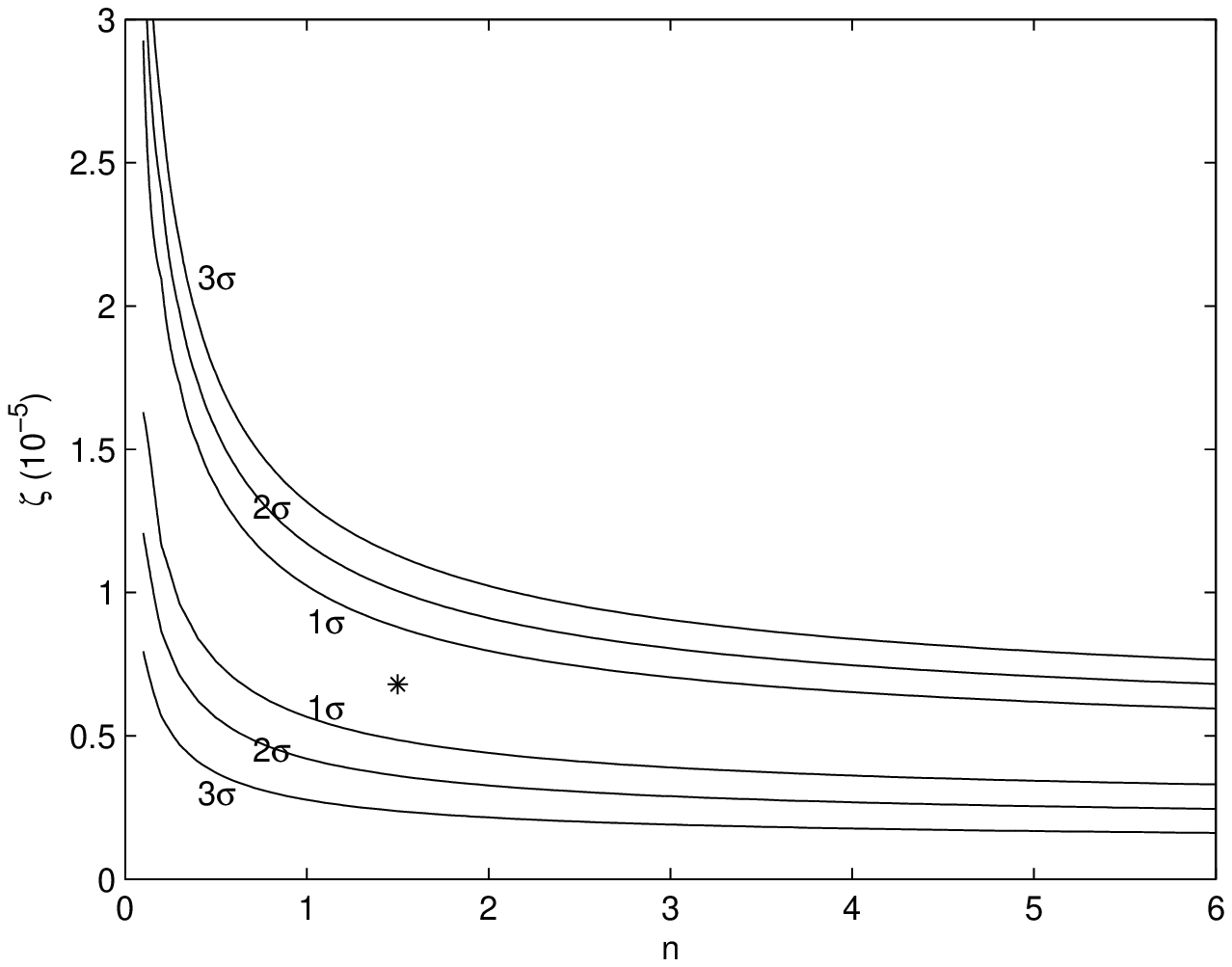,width=3.2truein,height=2.7truein}
 \psfig{figure=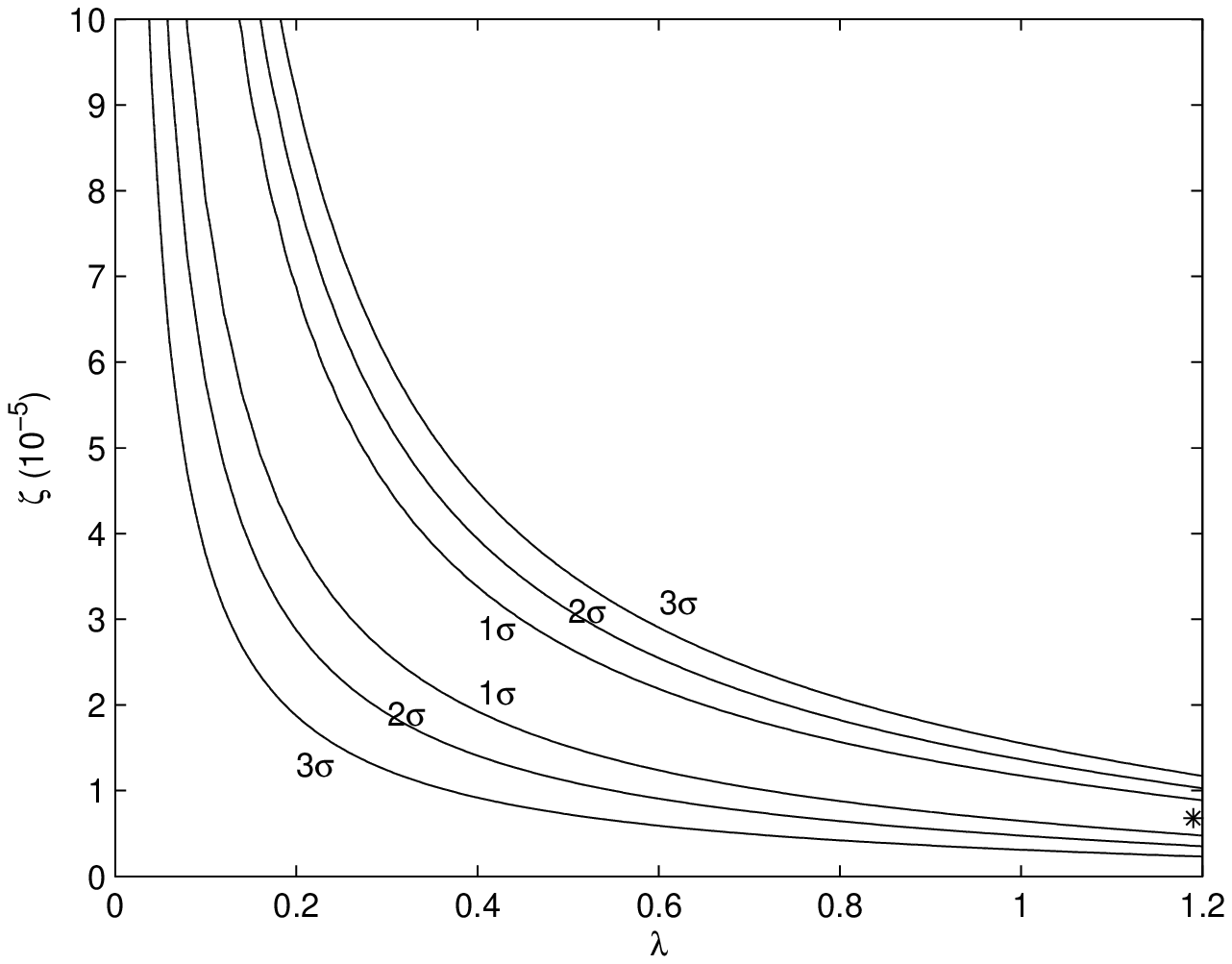,width=3.2truein,height=2.7truein}
\hskip 0.01in} \caption{$Left (Model I)$: The confident regions of
$(n,\zeta)$ obtained from KWM143 with a prior of
$\Omega_{m0}=0.275\pm0.016$. The best-fit results which is indicated
by the star are $(n,\zeta)=(1.5, 0.68\times10^{-5})$ with
$\chi^{2}_{min}=149.5672$. $Right (Model II)$: The confident regions
of $(\lambda,\zeta)$ obtained from KWM123 with a prior of
$\Omega_{m}$. The best-fit results which is indicated by the star
are $(\lambda,\zeta)=(1.2, 0.68\times10^{-5})$ with
$\chi^{2}_{min}=149.9395$.} \label{fig:KWM143_Om}
\end{figure*}

\begin{figure*}
\centerline
{\psfig{figure=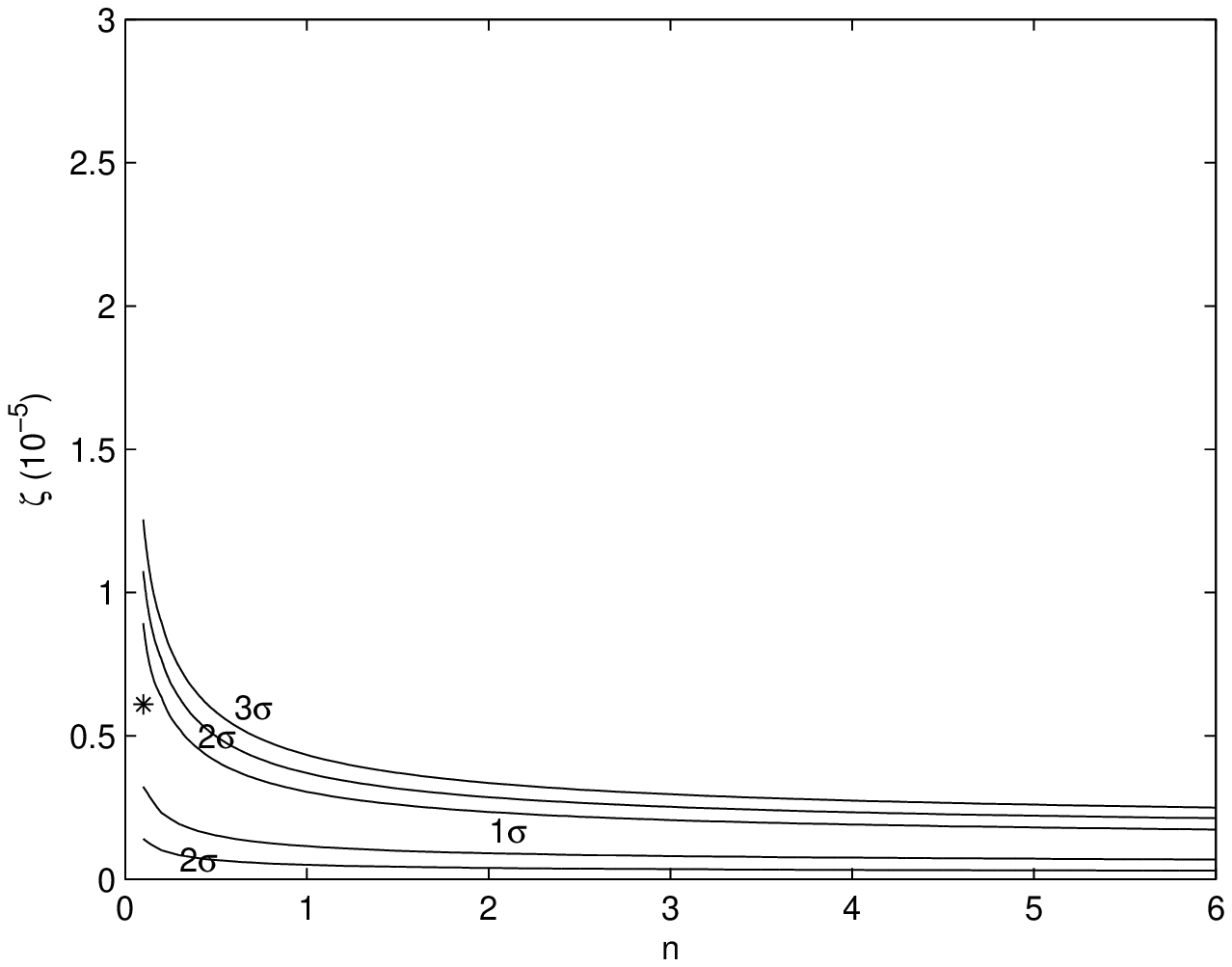,width=3.2truein,height=2.7truein}
 \psfig{figure=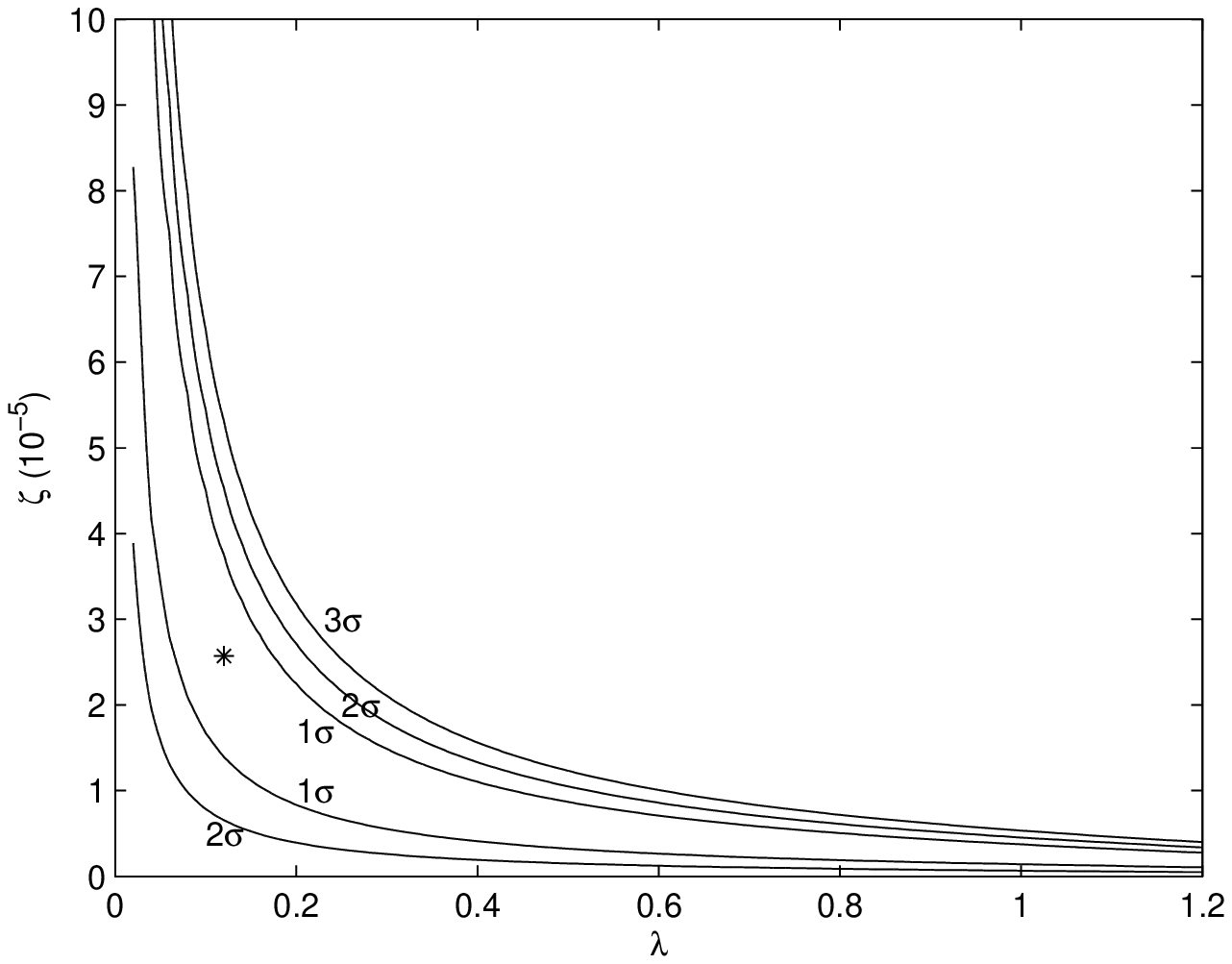,width=3.2truein,height=2.7truein}
\hskip 0.01in} \caption{$Left (Model I)$: The confident regions of
$(n,\zeta)$ obtained from KWM143+KWM23+VCS23 with a prior of
$\Omega_{m0}=0.275\pm0.016$. The best-fit results which is indicated
by the star are $(n,\zeta)=(0.1, 0.61\times10^{-5})$ with
$\chi^{2}_{min}=226.8156$. $Right (Model II)$: The confident regions
of $(\lambda,\zeta)$ obtained from KWM143+KWM23+VCS23 with a prior
of $\Omega_{m}$. The best-fit results which is indicated by the star
are $(\lambda,\zeta)=(0.12, 2.57\times10^{-5})$ with
$\chi^{2}_{min}=226.0966$.} \label{fig:total_Om}
\end{figure*}

\begin{figure*}
\centerline
{\psfig{figure=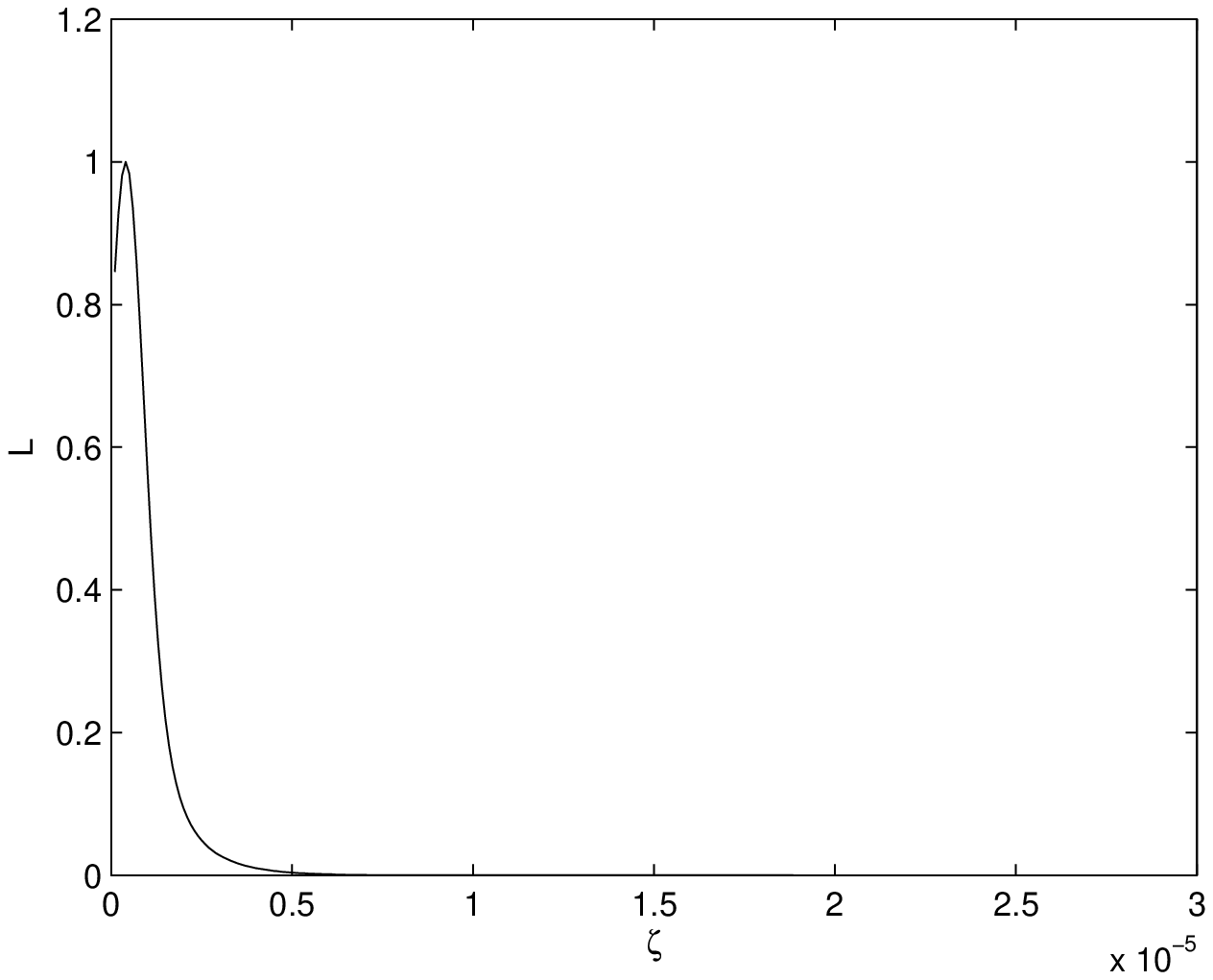,width=3.2truein,height=2.7truein}
 \psfig{figure=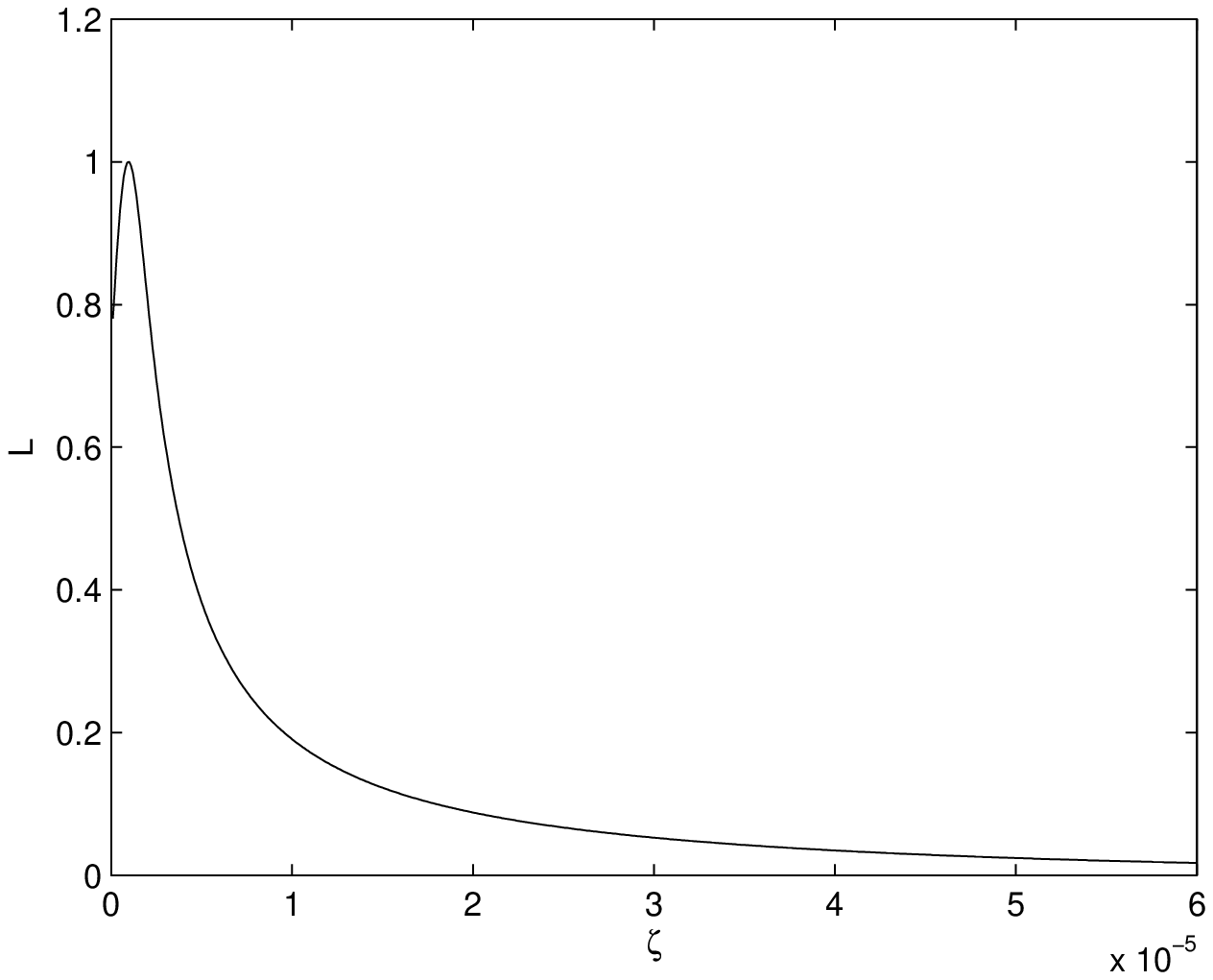,width=3.2truein,height=2.7truein}
\hskip 0.01in} \caption{$Left (Model I)$: The PDF of $\zeta$
obtained from VCS23. The most probable point is located at
$\zeta=0.04\times10^{-5}$. $Right (Model II)$: The PDF of $\zeta$
obtained from VCS23. The most probable point is located at
$\zeta=0.10\times10^{-5}$.} \label{fig:VCS23_L}
\end{figure*}

\begin{figure*}
\centerline
{\psfig{figure=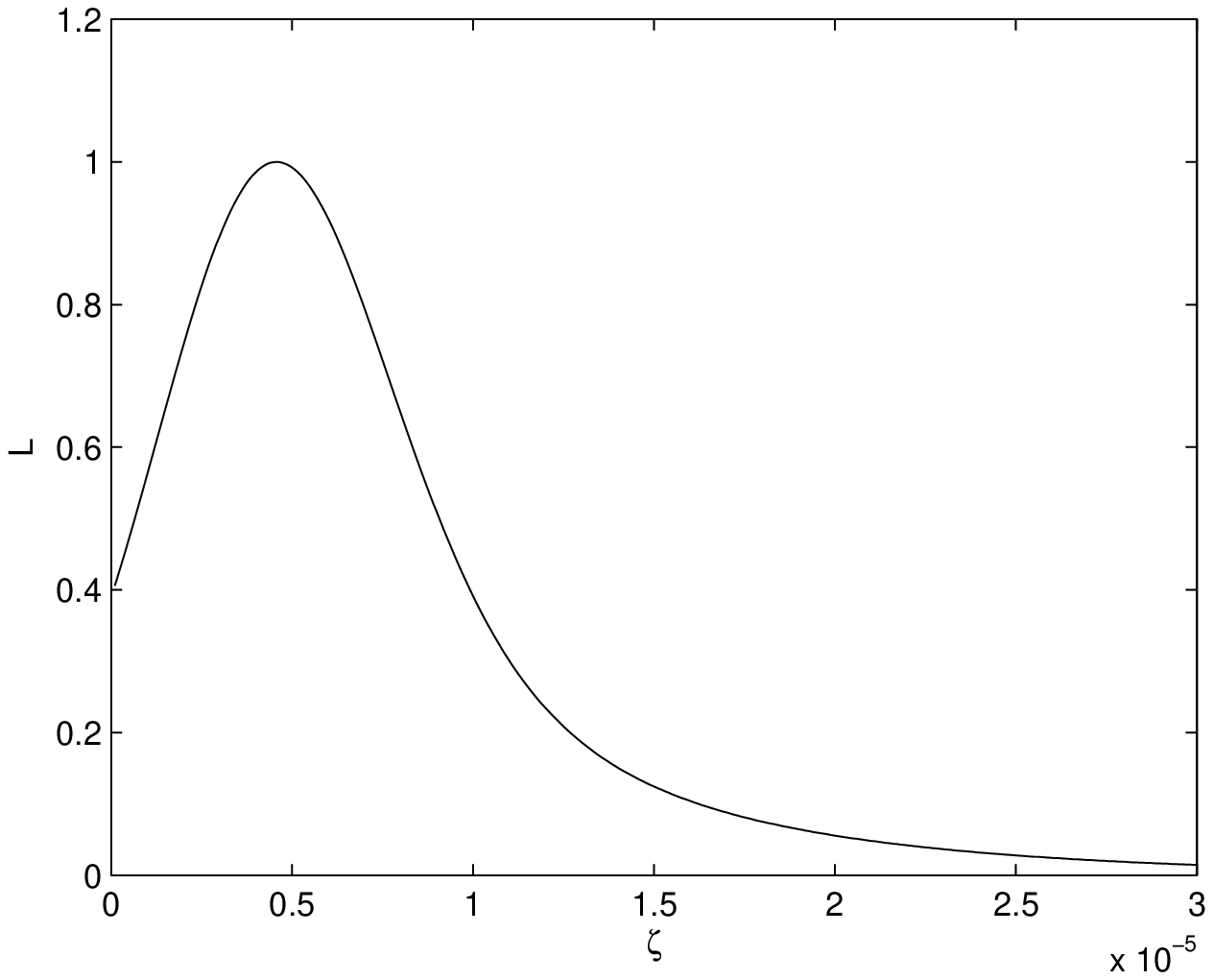,width=3.2truein,height=2.7truein}
 \psfig{figure=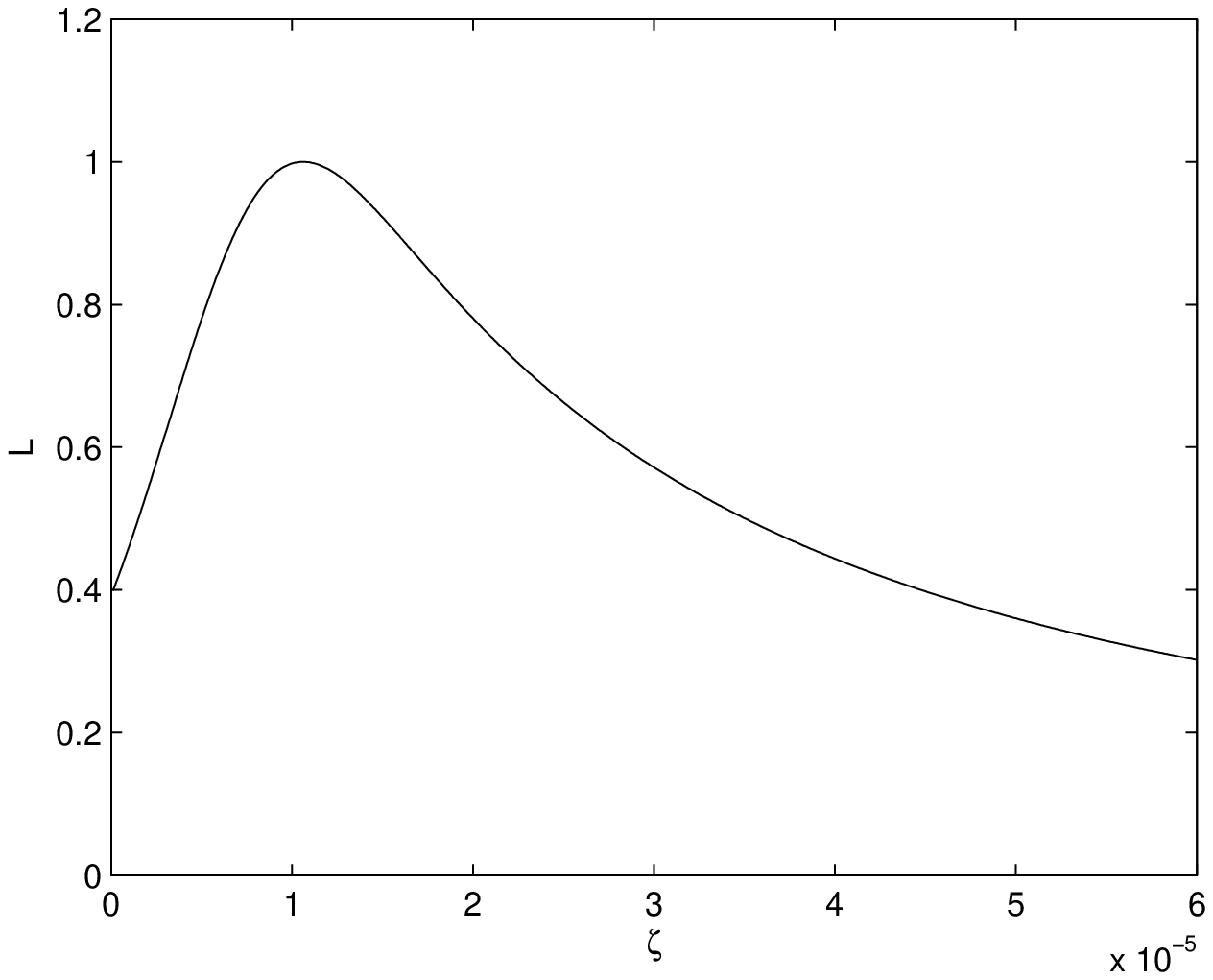,width=3.2truein,height=2.7truein}
\hskip 0.01in} \caption{$Left (Model I)$: The PDF of $\zeta$
obtained from VWM23. The most probable point is located at
$\zeta=0.46\times10^{-5}$. $Right (Model II)$: The PDF of $\zeta$
obtained from VWM23. The most probable point is located at
$\zeta=1.06\times10^{-5}$.} \label{fig:VWM23_L}
\end{figure*}

\begin{figure*}
\centerline
{\psfig{figure=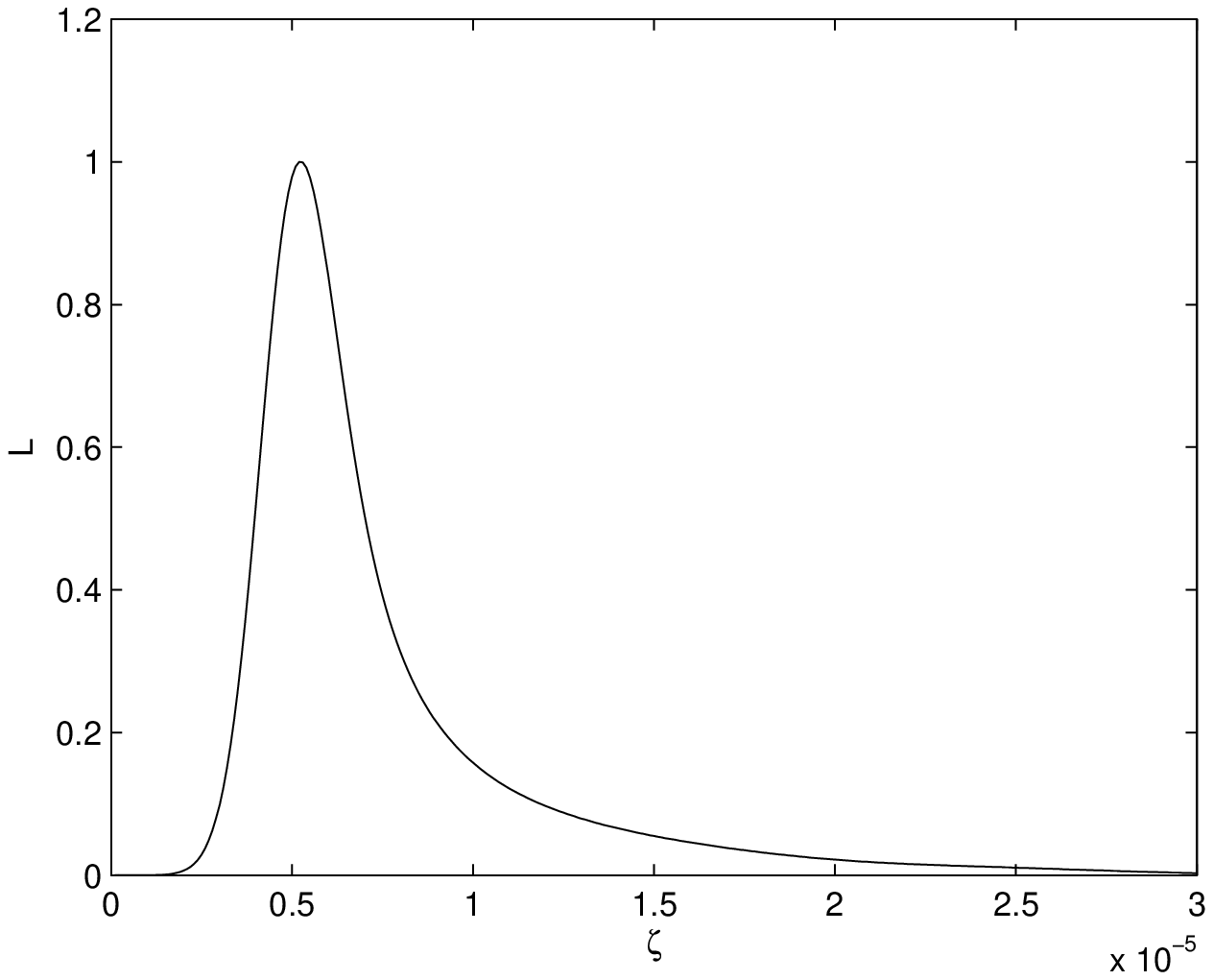,width=3.2truein,height=2.7truein}
 \psfig{figure=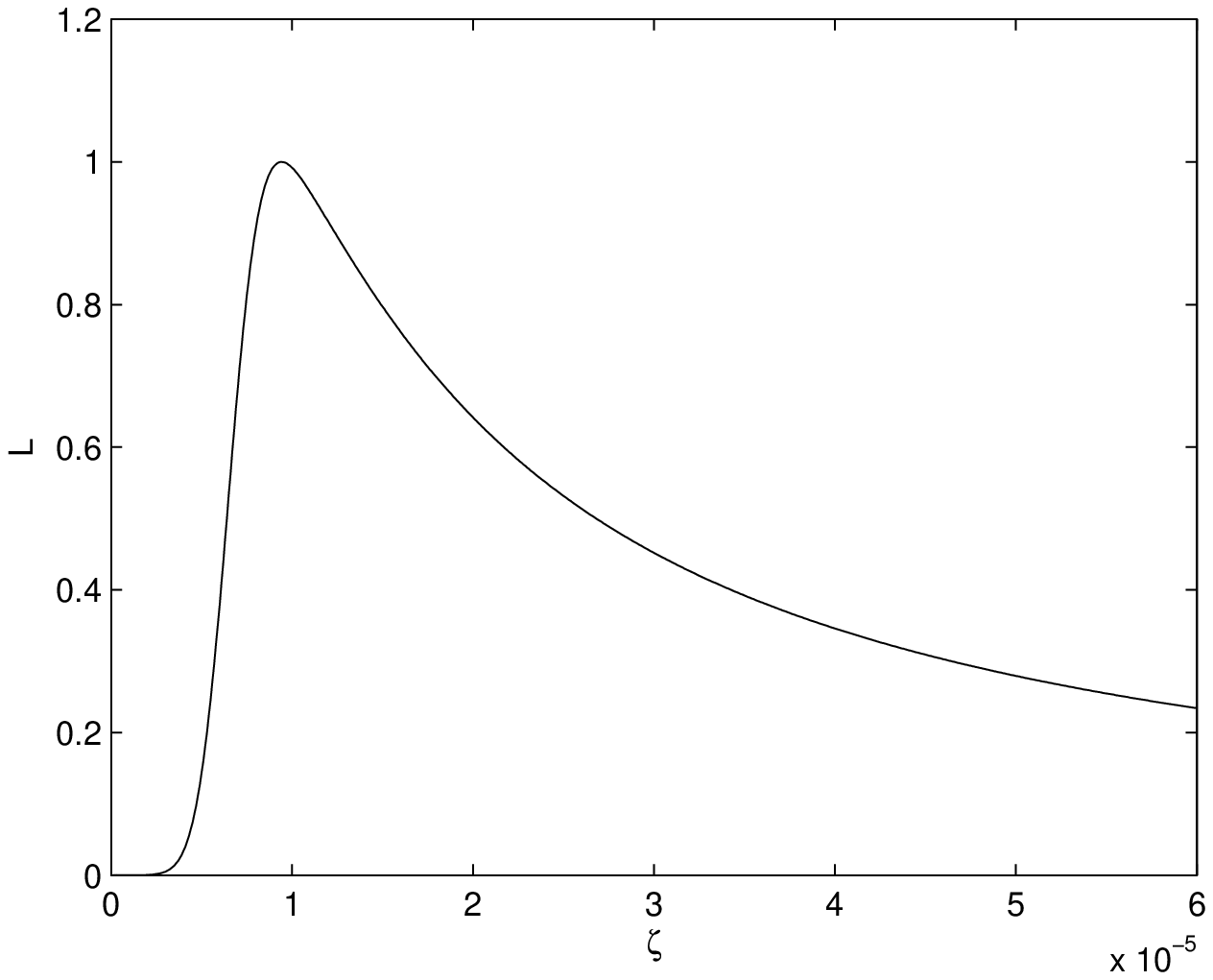,width=3.2truein,height=2.7truein}
\hskip 0.01in} \caption{$Left (Model I)$: The PDF of $\zeta$
obtained from KWM143. The most probable point is located at
$\zeta=0.52\times10^{-5}$. $Right (Model II)$: The PDF of $\zeta$
obtained from KWM143. The most probable point is located at
$\zeta=0.94\times10^{-5}$.} \label{fig:KWM143_L}
\end{figure*}

\begin{figure*}
\centerline
{\psfig{figure=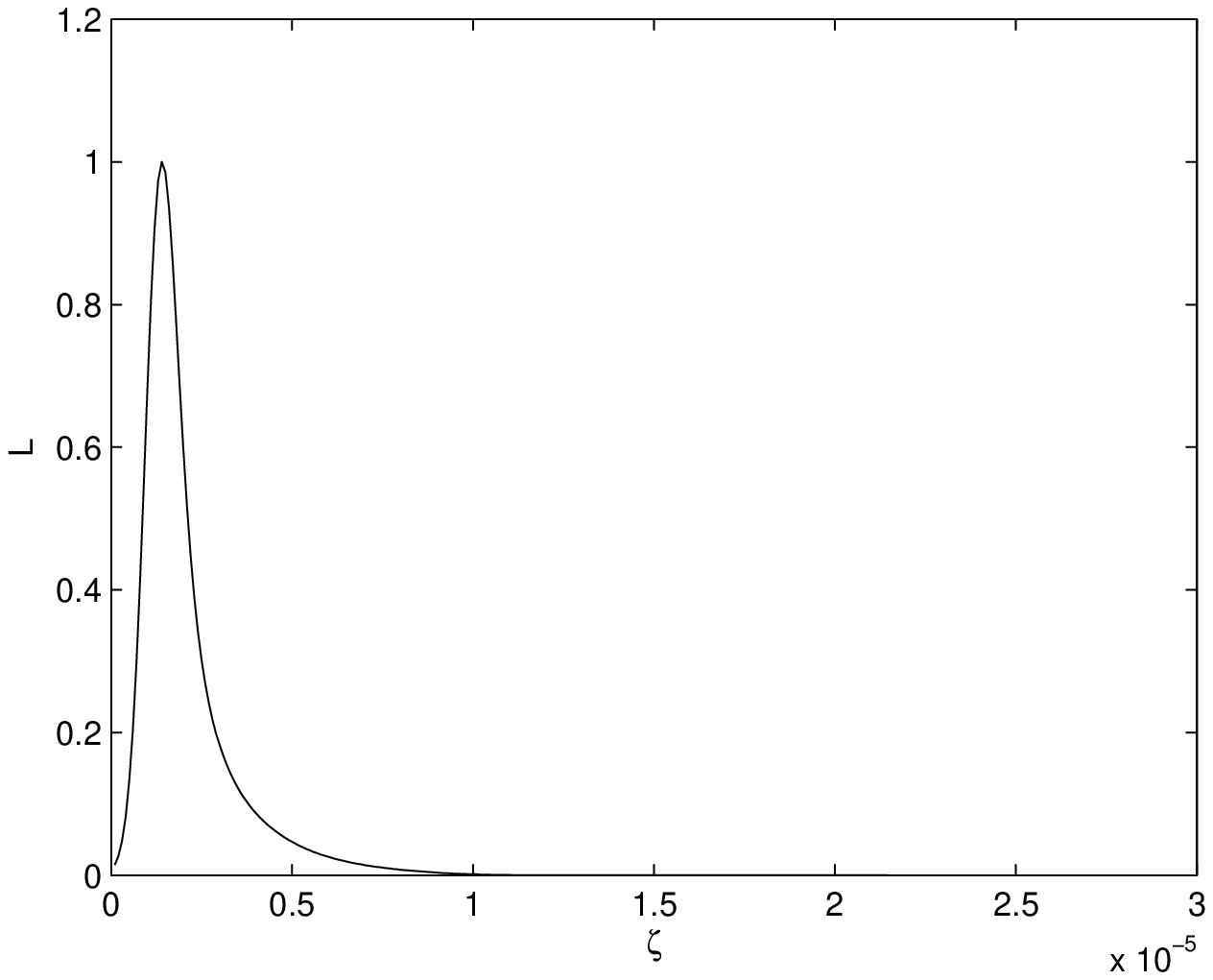,width=3.2truein,height=2.7truein}
 \psfig{figure=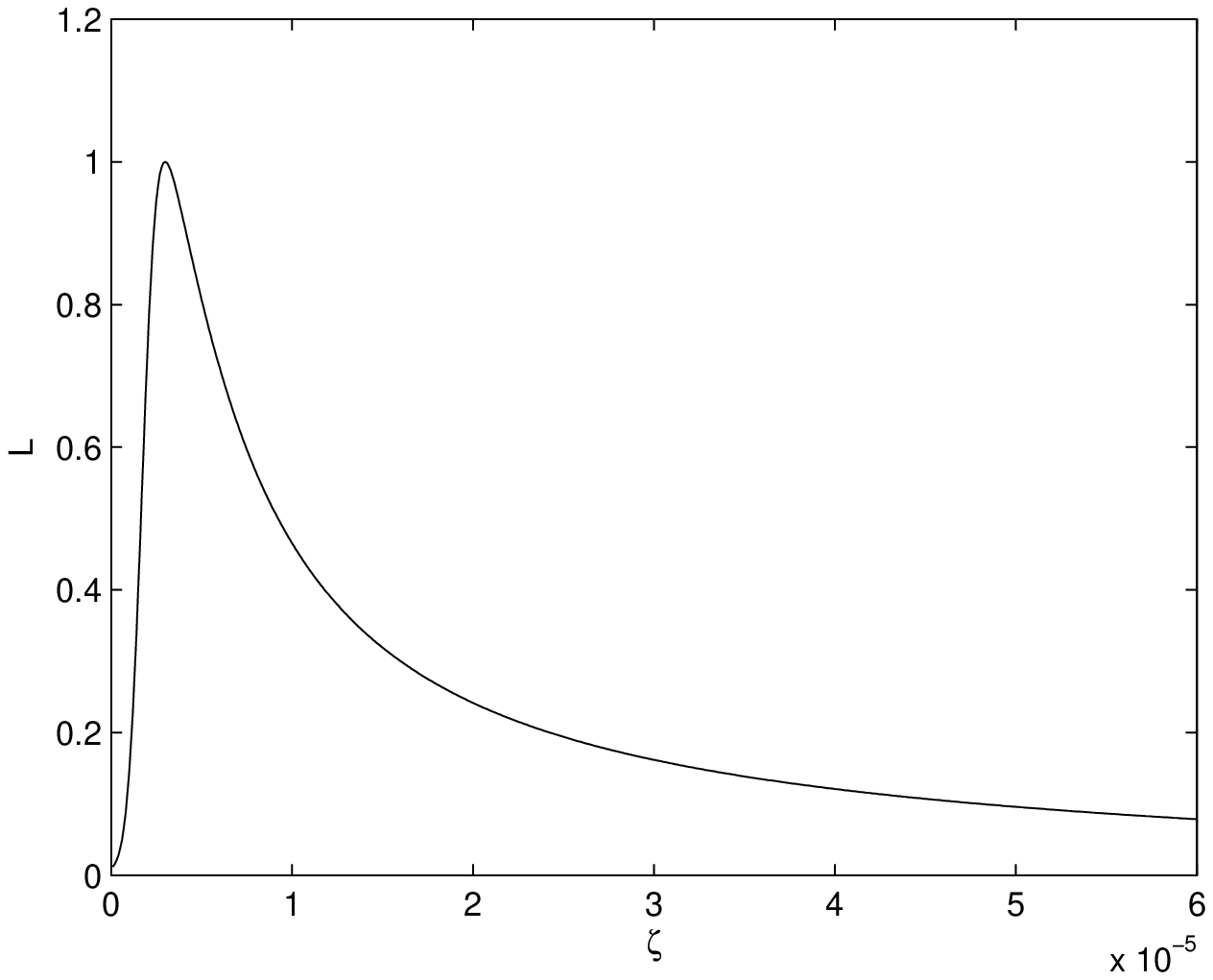,width=3.2truein,height=2.7truein}
\hskip 0.01in} \caption{$Left (Model I)$: The PDF of $\zeta$
obtained from KWM143+KWM23+VCS23. The most probable point is located
at $\zeta=0.14\times10^{-5}$. $Right (Model II)$: The PDF of $\zeta$
obtained from KWM143+KWM23+VCS23. The most probable point is located
at $\zeta=0.30\times10^{-5}$.} \label{fig:total_L}
\end{figure*}

\begin{figure*}
$\begin{array}{cc}
\includegraphics[width=0.5\textwidth,height=0.4\textwidth]{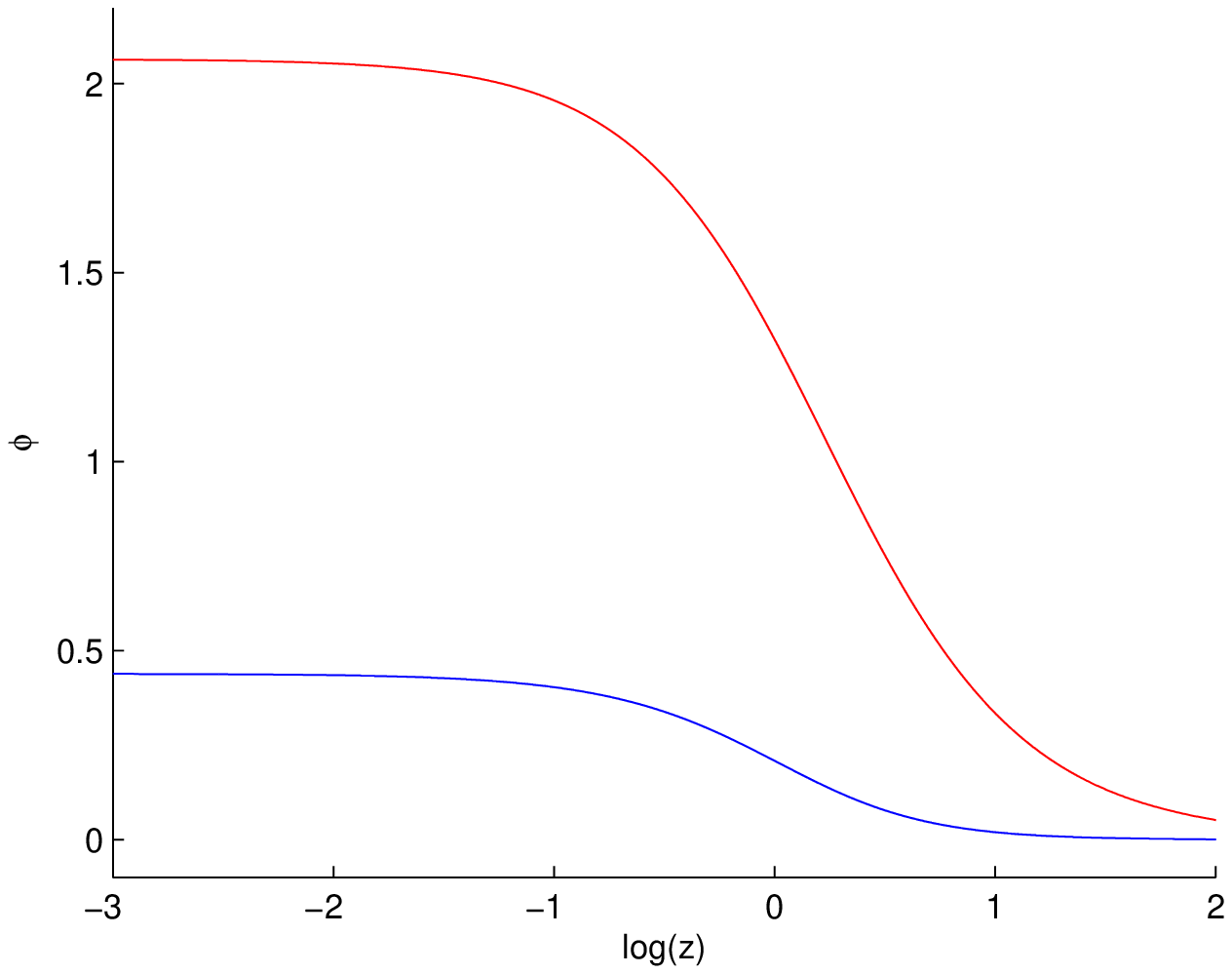} & \quad
\includegraphics[width=0.5\textwidth,height=0.4\textwidth]{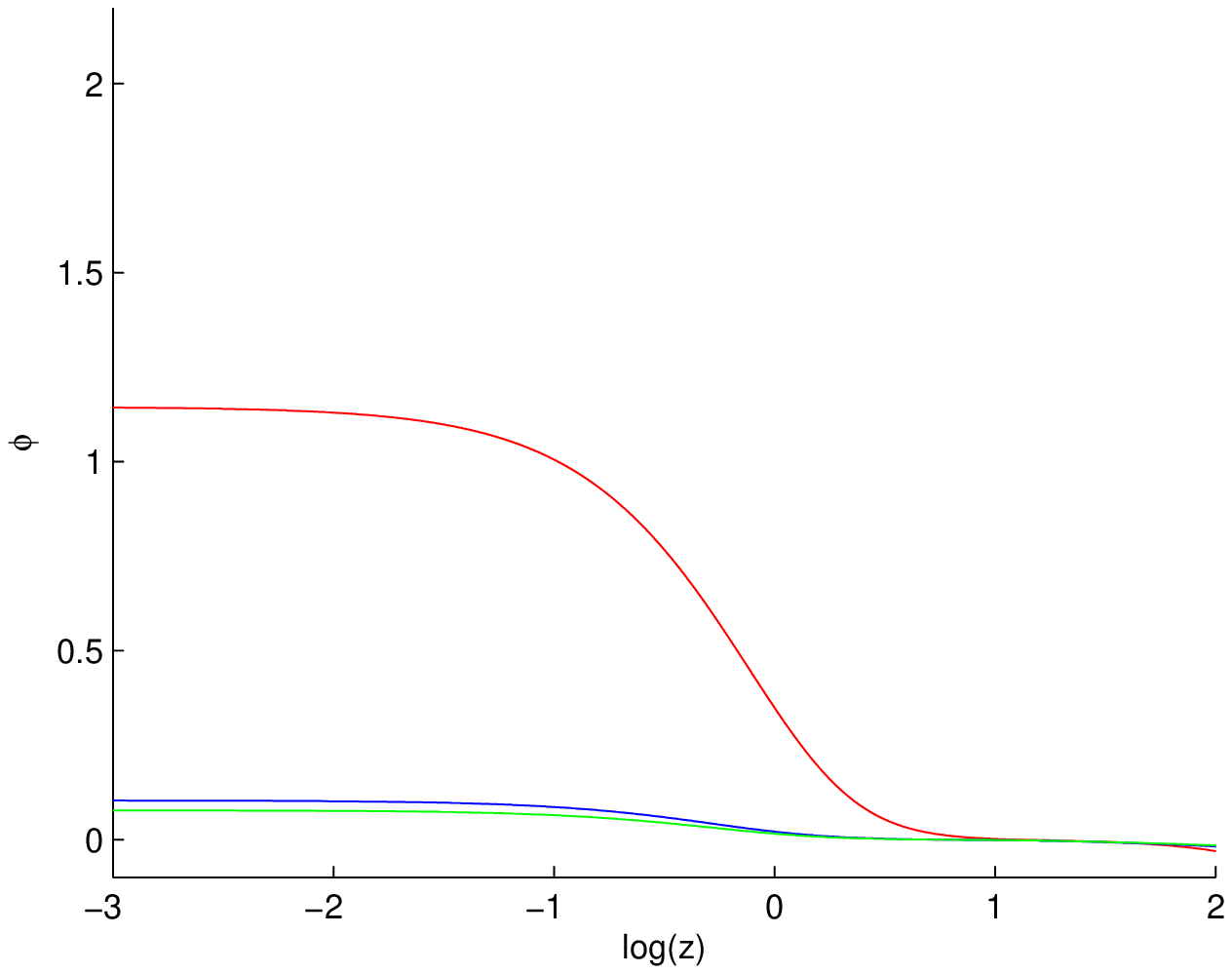} \\
\includegraphics[width=0.5\textwidth,height=0.4\textwidth]{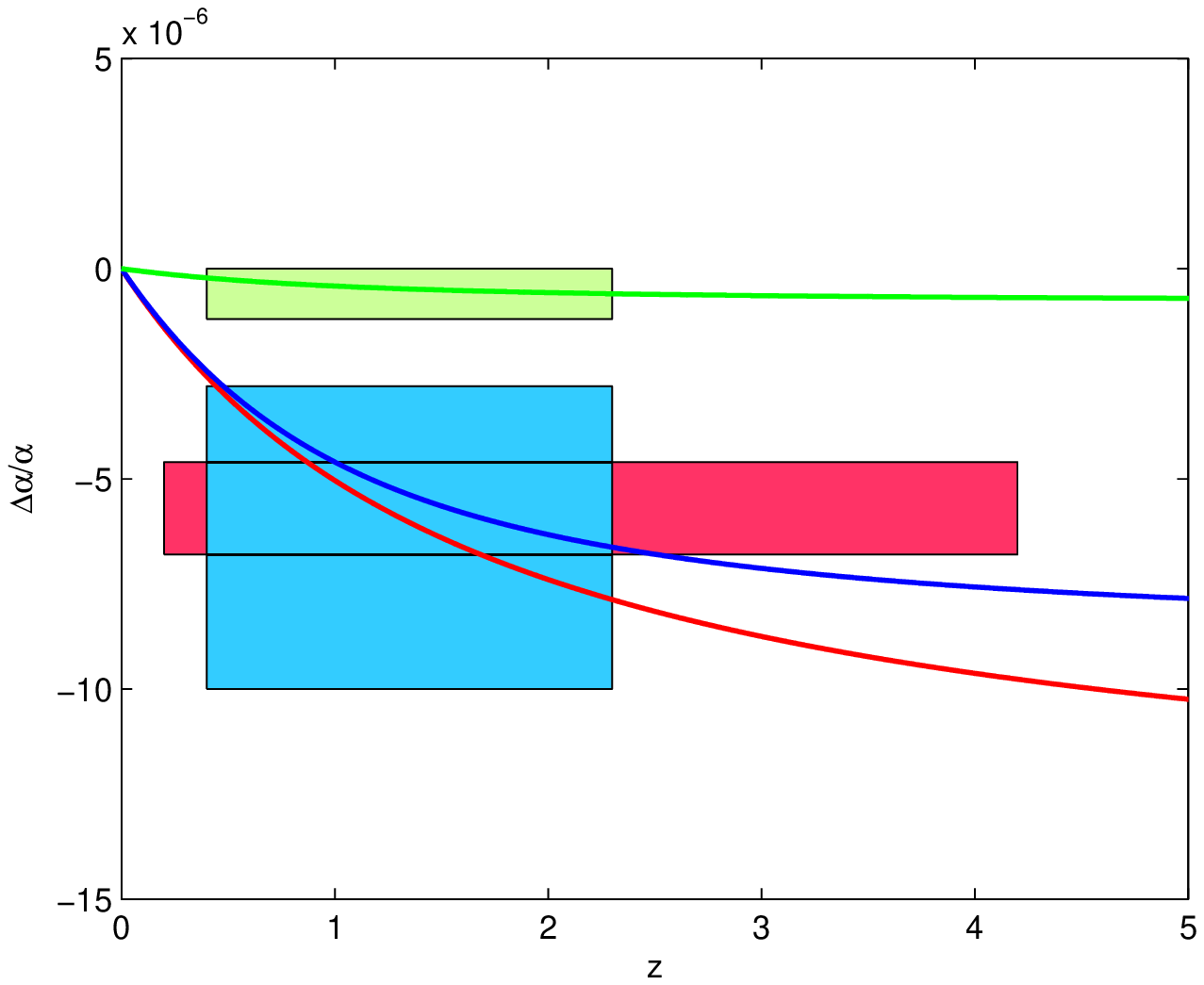} & \quad
\includegraphics[width=0.5\textwidth,height=0.4\textwidth]{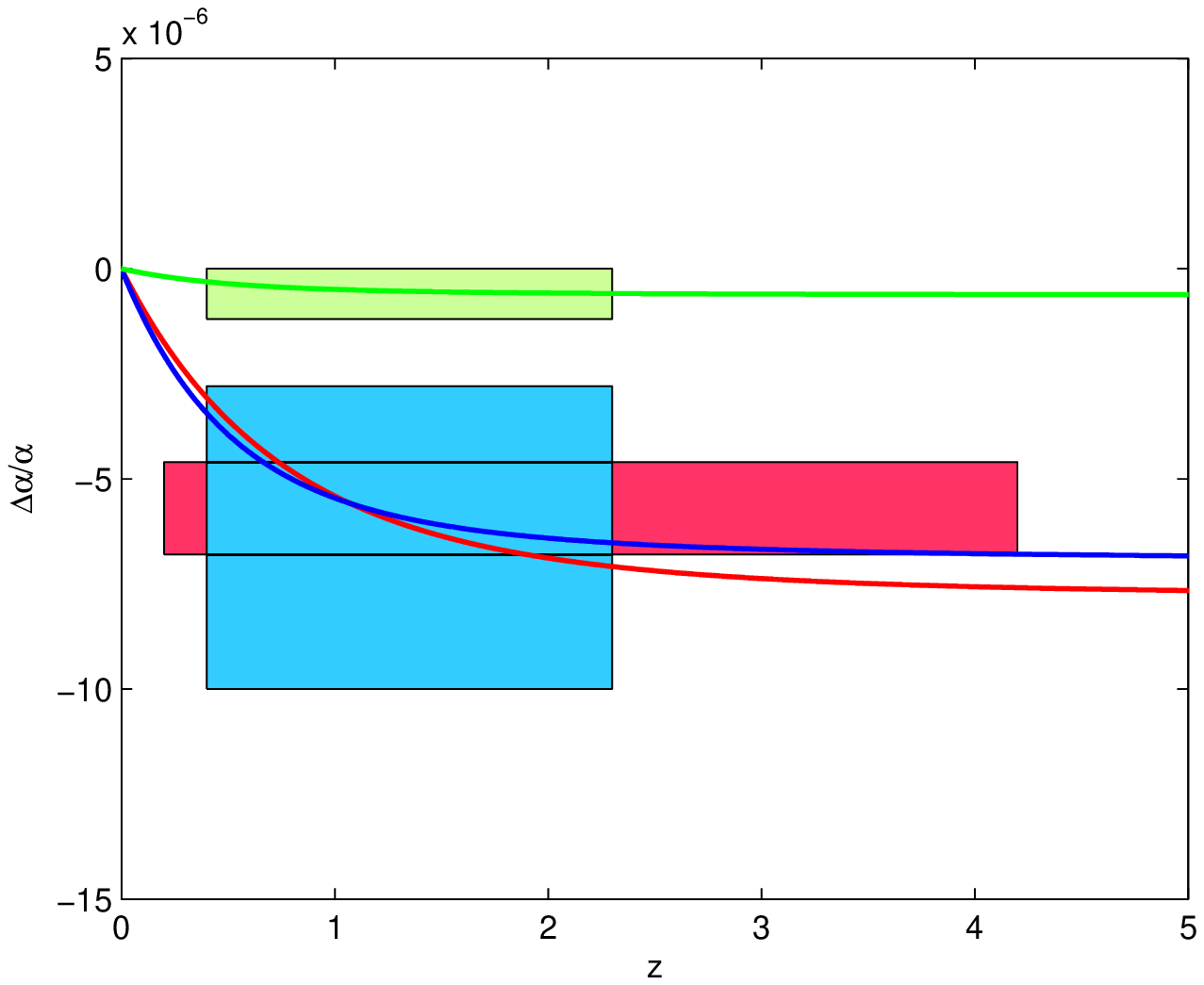} \\
\includegraphics[width=0.5\textwidth,height=0.4\textwidth]{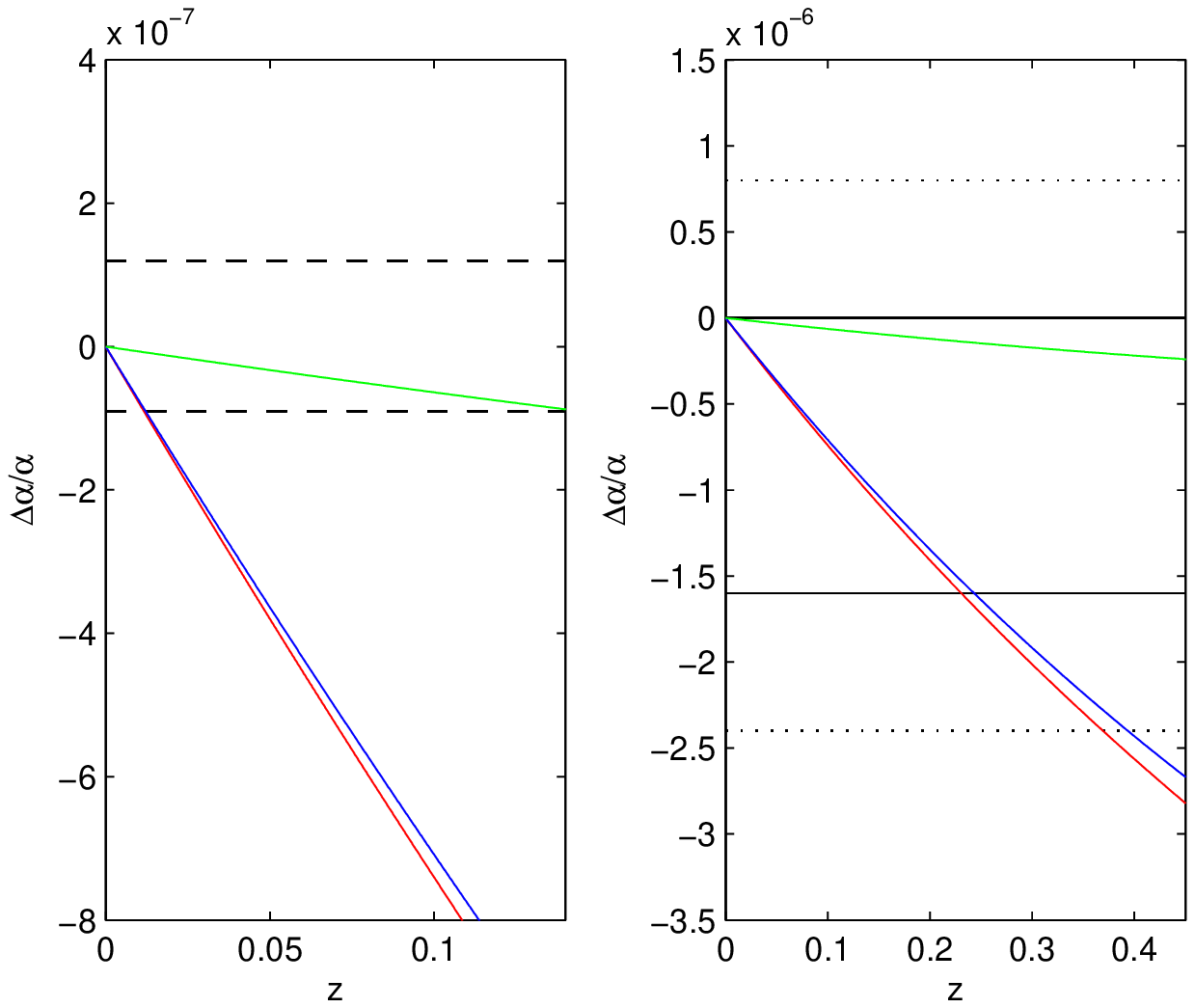} & \quad
\includegraphics[width=0.5\textwidth,height=0.4\textwidth]{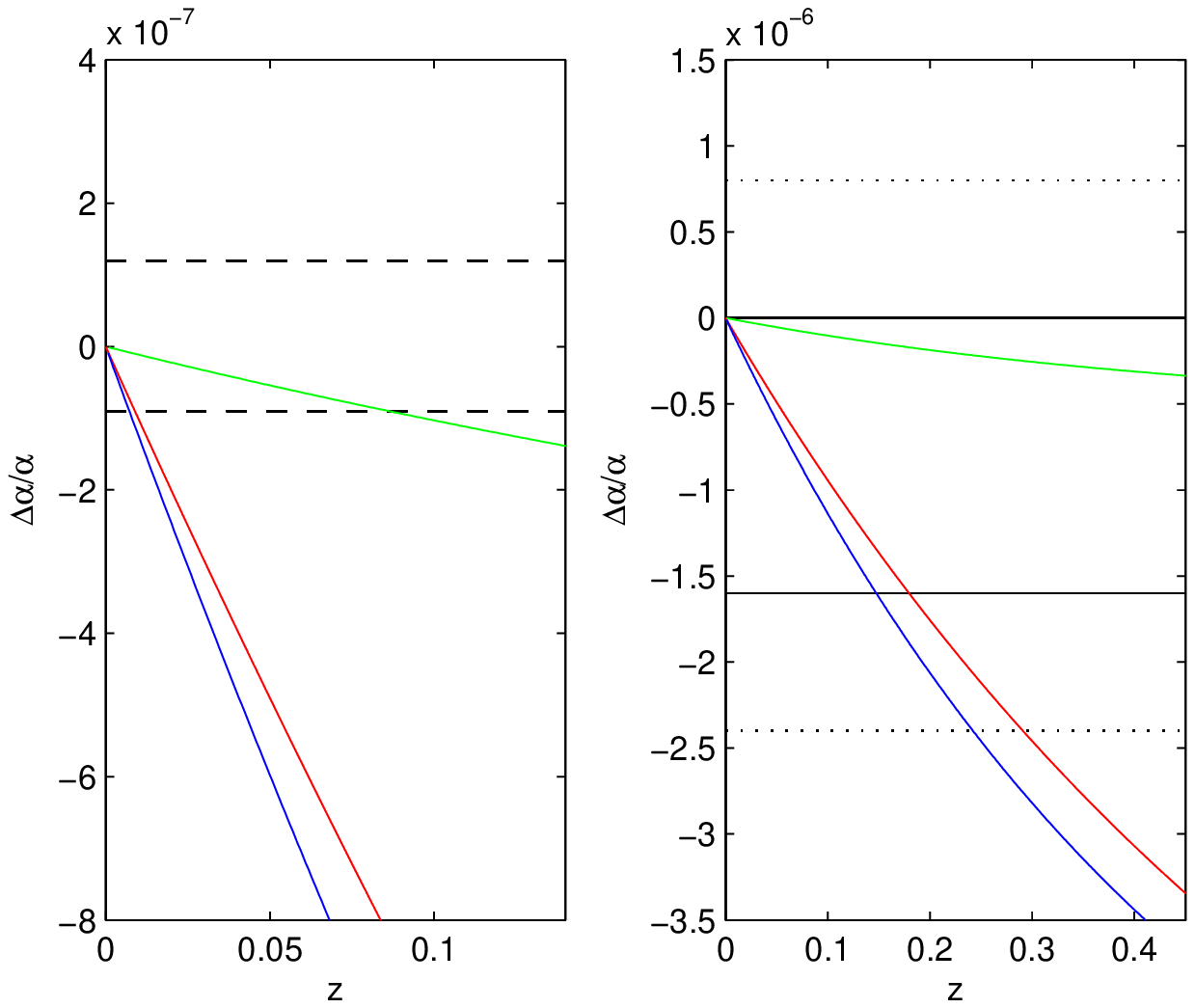} \\
\end{array}$
\caption{$Left$ $Top$: The evolution of the scalar field $\phi$
under the potential I with respect to redshift $z$, the red and blue
curves are obtained by the best-fit values of KWM143 and VCS23
(VWM23) respectively. $Left$ $Middle$: the evolution of
$\Delta\alpha/\alpha$ with respect to $z$. The red, blue and green
curves are obtained by the use of the best-fit values of KWM143,
VWM23 and VCS23 respectively, while the boxes are the corresponding
weighted values of QSO observations. $Left$ $Bottom$: The comparison
of the QSO results with Oklo bound (dashed lines) (left panel) and
meteorite bound (the solid and dotted lines correspond to 1$\sigma$
and 2$\sigma$, respectively) (right panel). $Right$: The same as the
$Left$ but for Model II.} \label{fig:phi_1}
\end{figure*}

In order to get the best-fit results of the parameters of the
Quintessence-$\alpha$ model, we apply the $\chi^{2}$ statistics to
the observational QSO data
\begin{equation}
  \chi^{2}(z;\Omega_{m0},n;\zeta)=\sum_{i}\left(\frac{(\Delta\alpha/\alpha)_{th,i}-(\Delta\alpha/\alpha)_{obs,i}}{\sigma_{i}}\right)^{2},
\end{equation}
where the subscripts "th" and "obs" stand for the theoretically
predicted value and observed ones respectively. In order to obtain
the purely results, we do not take into account other experimental
bounds as mentioned in Sec.\ref{sec:introduction} of our $\chi^{2}$
calculation, but the comparisons of the single QSO constraints with
other experiments are worth studying and are carried out in the
later sections. For the purpose of reducing the unnecessary
distractions arising from the intrinsic complexity of this scalar
field model, it is convenient to set reasonable priors on some of
the parameters.

As mentioned in the previous sections, one important discovery of
the cosmology is the present accelerated expansion. This discovery
indicates that the universe contains the so-called "dark energy"
component much more than the ordinary matter, i.e. in our
quintessence-$\alpha$ model, the parameter $\Omega_{m0}$ should
occupy a relative smaller proportion. Therefore, we adopt a Gaussian
distribution of $\Omega_{m0}=0.275\pm0.016$ as WMAP 7 suggests
(\cite{Komatsu..CMB}) to be a prior to constrain the
quintessence-$\alpha$ model. Thus the parameters which will be
constrained are $(n,\zeta)$ for Model I and $(\lambda,\zeta)$ for
Model II.

Our constraints results are shown in
FIG.\ref{fig:VCS23_Om}-FIG.\ref{fig:total_Om}. The corresponding $1\sigma$ errors of the parameters are summarized in Table.\ref{tab:results}.
Generally speaking, for Model I, the best-fit values of the Quintessence-$\alpha$ model
obtained from three data samples all favor a small value of $n$.
This feature is consistent with most other cosmic probes which show
that $n<1.5$ (\cite{Samushia..scalar}). This phenomena shows that the
scalar field evolves slowly in the universe. And the value of $n$ is
smaller, the scalar field model is closer to the standard
$\Lambda$CDM. Except that, the two contradictory samples VCS23 and
VWM23 both indicate $n=0.1$.
 The main differences between these constraints results
are the big discrepancies of the value of $\zeta$. The value
obtained by VWM23 is much larger than the other two samples, while
the VCS23 gives the smallest one.

Except that, from Table.\ref{tab:results}, one can see that the constraint of $\zeta$ in Model II
is much worse than Model I. Both the best-fit values of $\zeta$ and the $1\sigma$ upper bound are larger in Model II.
This can be seen as a signal of the different choices of the potential of the scalar field in Eq.\ref{Eq:V},
because the coupling strength between the electromagnetic field and the scalar field seems to be not as sensitive in Model II as
in Model I to the cosmological evolution.

For Model II, the constraints show similar trends of the parameters.
The best-fit value of $\lambda$ obtained by KWM143 is apparently
larger than the other two data samples. But the coupling constant
$\zeta$ of VCS23 is larger than KWM143 which is different from Model
I case.

Additionally, these results of two models are also consistent with
the Equivalence Principle test. But we should note that the
observations of the variations of fine structure constant can not
give efficient constraint on the cosmological parameter $n$ and
$\lambda$. Therefore it is difficult to identify the current
evolutionary state of the universe, i.e. the evolution of
$\Delta\alpha/\alpha$ is not as sensitive to the cosmological
parameters as to the coupling constant $\zeta$. One more point worth
noticing is that from our results, the variation of $\alpha$ can be
caused by large value of $n$ or $\lambda$, or the strong coupling
constant $\zeta$. This can be obtained from the comparisons of the
constraints, it is shown that the different weighted values of VWM23
and VCS23 is attributed to the variance of $\zeta$ instead of
cosmological parameters. However, the similar results of KWM23 and
VWM23 do not give constrict consistent constraint, VWM23 gives
smaller $n$ and $\lambda$ but larger $\zeta$. Therefore, the goal of
finding the reasons causing the variation of $\alpha$ is still
vague. But we should emphasize that the above conclusions are not
sure enough because of the insufficient constraints of $n$ and
$\lambda$ even the $1\sigma$ confident regions are not perfectly
obtained.

\subsection{the probability density function of $\zeta$}

The previous constraints show a 2 dimensional distribution of the
parameters $(n,\zeta)$ and $(\lambda,\zeta)$. In order to compare
the value of $\zeta$ with other tests, it is necessary to calculate
the probability density function (PDF) of $\zeta$ by marginalizing
the parameter $n$ or $\lambda$. Our results are presented in
FIG.\ref{fig:VCS23_L} to FIG.\ref{fig:total_L}.

Generally speaking, the results obtained are compatible with the
Equivalence Principle test which is $|\zeta|<5\times10^{-4}$. But
the differences between these calculations are also significant.
FIG.\ref{fig:VCS23_L} shows a apparent result that the best-fit
value of $\zeta$ is nearly zero for both models. The small coupling
constant indicates a case that the coupling between the
electromagnetic field and the scalar field is so weak that the fine
structure constant is nearly unchanged. While FIG.\ref{fig:VWM23_L}
and FIG.\ref{fig:KWM143_L} show different results. Both of VWM23 and
KWM143 give a similar value of $\zeta$ in each model. Their results
are consistent with each other and favor a variation of $\alpha$. It
is noteworthy that the constraint from VWM23 is not as strict as
KWM143 which may be attributed to the smaller size of this sample.
Comparing two models, the best-fit values of $\zeta$ are larger in
Model II than Model I, this implies that the coupling between the
scalar field and electromagnetic field is stronger in the
exponential potential than the inverse power law potential. Except
that, we also notice that once compared with the results obtained in
Sec.\ref{sec:gaussian}, an apparent discrepancy between the
constraints of $\zeta$ from VWM23 FIG.\ref{fig:VWM23_Om} and
FIG.\ref{fig:VWM23_L} emerges. This result is understandable because
the 2 dimensional constraints or the corresponding likelihood
function is non-gaussian. This comes from the quality of the data or
the non-linearity of the theoretical function.
\begin{table*}
\begin{center}\begin{tabular}{l||c|c||c|c}
 \hline
 & \multicolumn{2}{c||}{Model I}   & \multicolumn{2}{c}{Model II}  \\ \hline
 \normalsize               &  $\zeta$($10^{-5}$)        &  $n$           &   $\zeta$($10^{-5}$)      &  $\lambda$         \\ \hline
 \normalsize VCS23         &  (0, 0.10)                 &  (0, 3.4)      &   (0, 1.49)               &  (0.02, 0.30)       \\ \hline
 \normalsize VWM23         &  (0.08, 0.85)              &  (0, 3.4)      &   (0, 4.56)               &  (0.02, 0.5)        \\ \hline
 \normalsize KWM143        &  (0.33, 0.77)              &  (0, 3.5)      &   (0.53, 4.71)            &  (0.06, 0.54)       \\ \hline
 \normalsize Total         &  (0.06, 0.22)              &  (0, 3.4)      &   (0.08, 3.11)            &  (0.02, 0.34)       \\ \hline
\end{tabular}
\end{center}
\caption{The 1$\sigma$ confidence regions of the parameters of the two quintessence-$\alpha$ models. }\label{tab:results}
\end{table*}
\subsection{comparison with other experiments}

In this section, we consider the comparison of the QSO constraint
results with other experiments. In order to obtain a clear
impression, we plot the evolutions of the scalar field $\phi$ and
$\Delta\alpha/\alpha$ with respect to redshift $z$ in
FIG.\ref{fig:phi_1}(the top and middle panels). Firstly, we consider
the Oklo natural rector which provides a bound at 95\% confident
level,
\begin{equation}
  -0.9\times10^{-7}<\frac{\Delta\alpha}{\alpha}<1.2\times10^{-7}
\end{equation}
for $z=0.14$ \cite{Damour..oklo,Fujii1..oklo,Fujii2..oklo}. Except
that, the estimates of the age of iron meteorites at $z=0.45$
combined with a measurement of the Os/Re ratio resulting from the
radioactive decay $^{187}Re\rightarrow$ $^{187}Os$
gives\cite{Olive..oklo,Olive..ReOs,Fujii..ReOs}
\begin{equation}
  \frac{\Delta\alpha}{\alpha}=(-8\pm8)\times10^{-7}
\end{equation}
at $1\sigma$ and
\begin{equation}
  -24\times10^{-7}<\frac{\Delta\alpha}{\alpha}<8\times10^{-7}
\end{equation}
at $2\sigma$ \cite{Bento..ReOs}.

The results are presented in the bottom panels of
FIG.\ref{fig:phi_1}. Compared with the QSO results, the Oklo
measurements and meteorites estimates both favor an unchanged value
of the fine structure constant, because they are consistent with
VCS23 constraint except a tiny deviation of Model II bounded by the
Oklo measurement. Compared with Oklo, the meteorites observations
give a wider range of uncertainty, however, the VWM23 and KWM143
both violate this bound. Furthermore, the larger value of the slope
of Model II at low redshift shows a more drastic deflection from the
meteorites constraint. However, we should note that we discuss these
results only with the best-fit values of the parameters. Once the
uncertainties of the parameters are taken into account, the above
tendencies would be weakened. And the impressions of the
corresponding uncertainties can be achieved in the constraints as
FIG.\ref{fig:VCS23_Om} to FIG.\ref{fig:total_Om} show.

\section{discussions and conclusion}\label{conclusion}

In the present paper, we present the constraints of the cosmological
parameters on the Quintessence model by the measurements of the
variation of fine structure constant $\alpha$ from distant QSOs. By
the use of the Gaussian prior of $\Omega_{m0}$, three data samples
KWM143, VWM23, and VCS23 give apparent various constraints of the
parameters. For both of the two potential models, VCS23 shows the
smallest $\zeta$ which can be treated as an explanation of the
results of Ref. \cite{Chand..QSO} because the weak coupling derives a
weak interaction with the electromagnetic field. This leads to an
unchanging of $\alpha$. On the other hand, VWM23 and KWM143 present
a result that the values of $\zeta$ are larger especially the VWM23
one, while the constraints of $n$ for Model I and $\lambda$ for
Model II are different. And the strong coupling strength implies the
possibility of a variation of $\alpha$. In order to further study
this problem, we marginalize the cosmological parameter $n$ or
$\lambda$ and obtain the PDF of $\zeta$. The results confirm our
analysis that the VCS23 favors a nearly null result of $\zeta$.
Except that, the discrepancy between the VWM23 and KWM143 about
$\zeta$ disappeared and provide consistent constraints of it.
Combined the two models, we find that either a strong coupling or a
large value of the cosmological parameters (here refers to $n$ or
$\lambda$) can lead to an apparent variation of $\alpha$. Our
results show that the difference between VWM23 and VCS23 is caused
by the different coupling constant, while the similar results of
VWM23 and KWM143 have different reasons. The former is attributed to
a stronger coupling and the latter is caused by a different
evolution of quintessence scalar field. In order to obtain a
complete analysis, we also constrain the quintessence models with
all the available data as FIG.\ref{fig:total_Om} and
FIG.\ref{fig:total_L}. As a eclectic result, we find that the
coupling between the electromagnetic field and the scalar field is
stronger in Model II (the exponential potential) than in Model I.
This means that it is relatively easier for the inverse power law
potential to derive a change of the fine structure constant than the
exponential potential.

Furthermore, we should point out that the observations of the
variations of $\alpha$ by QSOs are not efficient as other cosmic
probes. This feature is reflected in their insensitive to the
cosmological parameters such as $n$ or $\lambda$, because the
sufficient constraints of the confident regions are also necessary
as the best-fit values.
From the confidence regions and the constraint errors of the parameters, we see that
the constraint of $n$ or $\Lambda$ is not as strict as other cosmic probes as supernovae or CMB \cite{Samushia..scalar}.
This is a relatively more obvious shortcoming of the observations of the variation of $\alpha$.
On the other hand, the supernovae observations do not give information about the coupling strength between the electromagnetic field and the
scalar field directly, so does its theoretical calculation. Therefore, the combination of the supernovae data and the $\Delta\alpha/\alpha$ data may provide us
more complete description of the universe. One possible way may be to use the supernovae data firstly and find the constraints of the parameters as $n$ or
$\lambda$, and then apply the results to constrain $\zeta$ and decide the coupling strength, because the mechanism of the supernovae is relatively more clearer
than the QSOs and thus the uncertainties of the cosmological parameters may be bound to be smaller.

Except that, further researches on the
distribution or the precise value of $\zeta$ are important. And the
comparisons between them with the QSO research are also imperative.
If we hope to get more
accurate description of $\Delta\alpha/\alpha$, measuring $\zeta$
accurately or combined with other observations will be
necessary (\cite{Amendola..sne}).
Moreover, the relationship between $\zeta$ and other quantity of the dark energy such as the equation of state is also meaningful \cite{Amendola..sne},
since the connection between the fundamental constant and the dark energy models can be indicated. Therefore the power of the observations
of the fundamental constants in studying the cosmic evolution could be searched more deeply.

$Note$ $added$, we notice that recently, the correlation of the
cosmic dipoles between the fine structure constant and the
supernovae are studied in Ref. \cite{Mariano..spatial}. From the
theoretical view, exploring this correlation under the scalar field
assumption and reconstruct the quintessence model is also worth
studying. And we will discuss this question in our future research.

\section{acknowledgement}
The authors would like to
thank Liu, W., Wang,  H., Chen, Y., Landau, S. J., Meng, X. and Ma,
C. for their helpful discussions and valuable suggestions. This
research is supported by the National Science Foundation of China
(Grant Nos. 11173006, 10773002, 10875012, 11175019), the Ministry of
Science and Technology National Basic Science program (project 973)
under grant No. 2012CB821804, and the Fundamental Research Funds for
Central Universities.


\end{document}